\documentclass[apj]{emulateapj}
\usepackage{amsmath}
\usepackage{amssymb}
\usepackage{epsfig}
\usepackage{apjfonts}

\newcommand{\cms}{\,{\rm cm$^{-2}$}\,}

\newcommand{\kms}{\,{\rm km\,s$^{-1}$}\,} 
\newcommand{\kmsmpc}{\,{\rm km\,s$^{-1}$\,Mpc$^{-1}$}\,}

\newcommand{\etal}{{ et~al.~}}

\newcommand{\ergs}{\,{\rm erg\,s$^{-1}$}\,}
\newcommand{\ergscm}{\,{\rm erg\,s$^{-1}$\,cm$^{-2}$}\,}
\newcommand{\Ms}{M_\odot}
\newcommand{\sfr}{\,M_\odot\,{\rm yr}^{-1}\,}

\newcommand{\mjysr}{\,{\rm MJy\,sr$^{-1}$}\,}
\shorttitle{Star Formation in Pavo}
\begin{document}


\title{A Multiwavelength View of Star Formation in Interacting Galaxies 
  in the Pavo Group }

\author{M. Machacek, M.L.N. Ashby, C. Jones, W. R. Forman}
\affil{Harvard-Smithsonian Center for Astrophysics \\ 
       60 Garden Street, Cambridge, MA 02138 USA} 
\author{N. Bastian}
\affil{Institute of Astronomy, University of Cambridge \\
    Madingley Road, Cambridge, CB3 0HA, UK}

\email{mmachacek@cfa.harvard.edu}

\begin{abstract}

 We combine {\it Spitzer} IRAC mid-infrared and {\it
  Chandra} X-ray observations of the dominant galaxies NGC\,6872 and
  NGC\,6876  in the Pavo group with archival optical and HI data to 
  study interaction-induced  star formation.
In the spiral galaxy NGC\,6872, 
$8.0\,\micron$ and $5.8\,\micron$ nonstellar emission having colors
  consistent with polycyclic aromatic hydrocarbons (PAHs) 
is concentrated primarily in 
clumps in three regions: in a $5$\,kpc radius outer ring about 
 the center of the spiral galaxy, in a bridge of emission connecting 
NGC\,6872's northern spiral arm to IC\,4970, and 
along the full extent of both NGC\,6872's
  tidal arms. PAH emission is 
  correlated with young star clusters and dense
  HI regions. We find no strong differences in 
the mid-infrared colors of the star-forming regions in the spiral
  galaxy NGC\,6872 as a function of position relative to the 
 tidally interacting companion galaxy IC\,4970. 
  We find eleven very luminous X-ray sources  
  ($\gtrsim (0.5 - 5) \times 10^{39}$\ergs) clustered to the
  southwest in NGC\,6872, near bright star-forming regions. In NGC\,6872's
  tidal features, young star clusters form at the boundaries of
  diffuse X-ray gas, suggesting that stars form as gas 
  stripped by the interactions cools. The nucleus of 
 NGC\,6872 is a weak X-ray point source with $0.5-8$\,keV luminosity of 
 $8.5 \times 10^{39}$\ergs, but there is little evidence in the inner
 $1$\,kpc of NGC\,6872  for PAH emission from recent star formation or
 nuclear activity. However, a $4$\,kpc `stream', leading from the
 outer ring of NGC\,6872 to the nucleus, may signal the transport of 
 interstellar matter into NGC\,6872's nuclear region. 
 Nonstellar emission, 
 consistent with PAH emission, is also found in the central region
 of elliptical galaxy NGC\,6877, companion to the dominant Pavo group 
 elliptical galaxy NGC\,6876. However, in the central region of NGC\,6876, 
 the dust emission is more likely due to silicate emission from old AGB stars. 
 
\end{abstract}

\keywords{galaxies: clusters: general -- galaxies:individual 
(NGC\,6872, NGC\,6876, NGC\,6877) -- galaxies: interactions --
  infrared: galaxies -- X-rays: galaxies}

\section{Introduction}
\label{sec:introduction}

Galaxy collisions and mergers have long been known to influence
star formation in galaxies 
(see, e.g., the review by  Struck 1999 and references therein) 
and likely play a fundamental role in the
transformation of the population of gas-rich spiral galaxies at high 
redshift into the spheroidal and elliptical galaxies that 
dominate dense galaxy groups and clusters at the present epoch. Probing the 
details of the gravitational and hydrodynamical processes that affect
star formation during these collisions has only recently become
possible with the advent of high spatial resolution observations
 that are capable of resolving star-forming complexes within nearby
galaxies across the full electromagnetic spectrum (radio to X-ray), 
and that allow us to measure the thermodynamic properties of cold
and hot gas in and around these galaxies.

Early studies of star formation in interacting galaxies were 
based primarily on observations in the optical and far-infrared 
wavebands.  
These studies focused  
on global star formation properties and
correlations between observable star formation indicators. 
For example, Kennicutt \etal (1987) found that 
H$\alpha$ emission and far-infrared to blue luminosity 
ratios were enhanced in interacting galaxies compared to galaxies in
non-interacting samples. More recent optical studies using large area redshift
surveys suggest H$\alpha$ equivalent widths may be correlated with 
galaxy pair separation
(e.g., Barton \etal 2000 using the second  CfA redshift survey; 
Lambas \etal 2003 using the 2dF survey; Nikolic, Cullen \& Alexander
2004 using the SDSS survey).  However, extinction in the
optical wavebands and dust heating by evolved AGB stars in the
far-infrared made global star formation activity in galaxies 
difficult to measure, and contributed to the large scatter in 
these relationships. Studies of individual star-forming 
regions within interacting galaxies were largely limited to optical wavebands, 
because of the low spatial resolution of the far-infrared data.
From these optical studies, the spatial distributions and masses of star
forming clumps were found to be diverse. While star formation was
often enhanced in the nuclear region of the dominant galaxy of an  
interacting galaxy pair, star formation was also prevalent at large radii in 
extended  tidal features. Masses of these star-forming clumps 
span more than six orders in magnitude, from young ($2-6$\,Myr old)
$\sim10^2-10^3\Ms$ star associations, that may rapidly dissolve 
(Werk \etal 2008), to bound $\sim 10^5-10^6\Ms$ systems, with masses similar
to globular clusters (Trancho \etal 2007), and $10^7 - 10^8\Ms$
systems, sometimes called tidal dwarf galaxies. With the launch of 
the Spitzer Space Telescope
(Werner \etal 2004), high spatial resolution mid-infrared 
imaging and spectroscopy of star-forming regions within nearby
galaxies became possible. Using mid-infrared luminosities and colors,  
nonstellar emission from polycyclic aromatic hydrocarbon (PAH)
molecules, excited  by newly formed stars in dusty regions,  
could be separated from the diffuse starlight and the 
silicate emission expected from evolved stars, thus completing
the census of global star formation in the galaxy. With the high
angular resolution of the {\it Spitzer} Infra-Red Array Camera 
(IRAC; Fazio \etal 2004),  subtle 
differences in the star-forming clumps could be probed as a function
of their spatial distribution within the interacting system.

Most recent mid-infrared studies of  interaction-induced star formation have 
focused on interacting galaxy pairs in isolation from their
environment.  Galaxies in the largest such study, the Spitzer Spirals, 
Bridges and Tails Interacting Galaxy Survey (SSB\&T; Smith \etal 2007), were 
optically selected from the Arp catalog of interacting galaxies (Arp 1966) 
to include only nearby, isolated galaxy pairs that exhibited tidal
features, and specifically excluded triple and higher multiple galaxy 
systems and merger remnants. Smith \etal (2007) compared the mid-infrared 
properties of interacting galaxies in the SSB\&T sample,  
as a whole, to those of normal, noninteracting spiral 
galaxies drawn from the Spitzer Infrared Nearby Galaxy Survey
(SINGS; Kennicutt \etal 2003). On average, they found that 
interacting galaxy pairs have redder mid-infrared colors, and that star 
formation is enhanced and more centrally concentrated in the
dominant spiral galaxy of the interacting pair than in normal spiral 
galaxies. No evidence was found
for a correlation between mid-infrared colors and galaxy pair
separation in the interacting galaxies, as might have been expected 
from the optical data. 
Also, no strong differences were found between the mid-infrared 
colors in the interacting galaxies' stellar disks and those measured
in their tidal bridges and tails. However, since the study averaged 
over interactions with different orbital characteristics and galaxy 
masses, subtle differences related to the details of the interaction 
could be washed out. Individual case studies of a handful of isolated, 
interacting galaxy pairs, using ultraviolet, optical and
mid-infrared data, suggested that interaction-induced star formation 
occurs in clumpy bursts. The ages and spatial distribution of the 
star-forming clumps may reflect the stage and orbital parameters of
the collision, as inferred from numerical simulations of the collision 
(e.g. Wang \etal 2004 for NGC4038/4039; Smith \etal 2005a for Arp
107; Elmegreen \etal 2006 for NGC\,2207; Hancock \etal 2007 for Arp 82). 

Results from the DEEP2 galaxy survey show that the fraction of blue, 
star-forming galaxies is rapidly changing between redshifts 
$1.3 \gtrsim z \gtrsim 1$, and that galaxy evolution at these redshifts
occurs not in isolated galaxy pairs, but predominantly in 
moderately massive galaxy groups with $\lesssim 10$ galaxy members 
and velocity dispersions $200 \lesssim \sigma_v \lesssim 400$\kms 
(Gerke \etal 2005; Cooper \etal 2006). Thus understanding how 
the group environment affects star formation in interacting 
galaxies is vital to  understanding how dusty, blue star-forming disk galaxies
evolve into the early-type galaxies, hosting little or no ongoing star 
 formation, that dominate groups and rich clusters today. By including X-ray 
observations with the other traditional (optical and infrared)
tracers of star formation activity, we have direct observational
constraints on the temperature and density of the intragroup gas 
(IGM) surrounding these galaxies, and on the three-dimensional motions 
of the interacting galaxies with respect to the group IGM 
(see, e.g. Merrifield 1998; Vikhlinin \etal 2001; Machacek \etal 2005a,
2005b). X-ray observations also help identify nuclear activity and hot gas 
flows that might trigger or quench subsequent star formation.

\begin{figure}[t] 
\begin{center}
\includegraphics[height=2.37in,width=3in]{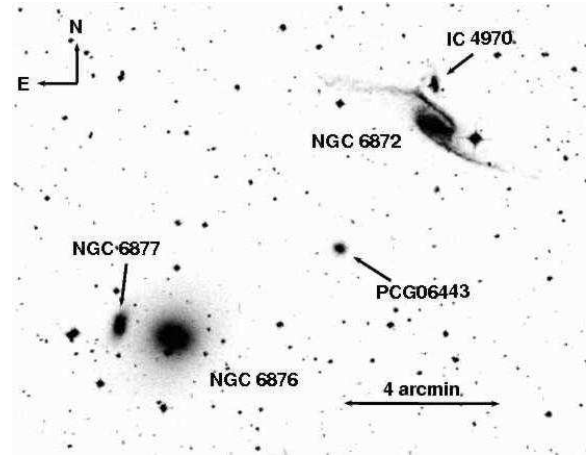}
\caption{ {\small DSS Bj-band image, taken with the UK $48$\,inch Schmidt
  Telescope, of galaxies near the core of the Pavo group.}
 }
\label{fig:pavodssgrp}
\end{center}
\end{figure}
The Pavo galaxy group is an important nearby laboratory to study the 
evolution of galaxies within their group environment, because it  
shares many properties with the high redshift galaxy groups found in
the DEEP2 galaxy survey. 
Like those galaxy groups, the Pavo group is moderately massive with a 
velocity dispersion, derived from its $13$ member galaxies, 
of $\sim 425$\kms, and 
contains a gas-rich spiral galaxy
(NGC\,6872) interacting with its environment. 
Yet, at $55.5$\,Mpc, the Pavo group is close enough for 
star-forming clumps with typical separations of $\sim 1$\,kpc (Elmegreen \&
Efremov 1996) to be resolved in {\it Chandra} X-ray and {\it Spitzer} 
mid-infrared images, as well as in the optical bands. 

The Pavo group is dynamically young. It hosts a variety 
of galaxy and group interactions expected to be important at high 
redshift.  In Figure \ref{fig:pavodssgrp} we show a Bj band image of the two
brightest galaxies in the Pavo group: the elliptical 
galaxy NGC\,6876 at the center of the group and the large, tidally
distorted SAB(rs)c spiral galaxy NGC\,6872 located $8\farcm7$ to the 
northwest. The large spiral galaxy, NGC\,6872, and its spheroidal companion
galaxy, IC\,4970,  located $1\farcm12$ to the north near a break in 
NGC\,6872's northern tidal arm, have long been known to be a tidally 
interacting galaxy pair (Vorontsov-Velyaminov 1959), and are  
well-studied in the optical (Mihos \etal 1993; Bastian \etal 2005) and 
radio (Horellou \& Booth 1997; Horellou \& Koribalski 2003; 2007) 
wavebands. The stellar bridge pointing from the northern tidal arm of 
NGC\,6872 to its less massive companion IC\,4970 indicates the tidal 
interactions between these two galaxies are ongoing. The $74$\kms 
radial velocity difference\footnote{Line-of-sight velocities 
and redshifts are taken from the CfA  Redshift Survey by Martimbeau \& Huchra. 
Online versions of the catalog, supporting software, and documentation 
are available at http://cfa-www.harvard.edu/$\sim$huchra/zcat.} between 
NGC\,6872 and IC\,4970 suggest that these galaxies form a spiral
dominated subgroup. {\it XMM-Newton} X-ray observations revealed a 
$90$\,kpc long X-ray trail of hot ($\sim 1$\,keV) gas linking, in
projection, NGC\,6872/IC\,4970 to NGC\,6876, the dominant Pavo group
elliptical galaxy. The properties of the 
X-ray gas in the trail, coupled with the large 
$850$\kms radial velocity difference between NGC\,6872 and NGC\,6876, 
provide dramatic evidence for the highly supersonic passage of 
the NGC\,6872/IC\,4970 subgroup through the dense $\sim 0.5$\,keV IGM gas 
in the Pavo group core (Machacek \etal 2005b). Although difficult to 
identify today, such interactions would likely have been frequent at 
high redshift, when galaxy groups and clusters were forming at the 
intersections of large scale filaments. 

The dominant elliptical galaxy, NGC\,6876,  
also shows evidence for recent interactions. Hubble Space Telescope
observations of the inner $1''$ of  
NGC\,6876 show a depressed central surface brightness distribution 
surrounded by a stellar torus, suggesting the presence of a binary
black hole from a previous merger (Lauer \etal 2002). The companion 
elliptical galaxy, NGC\,6877, ($1\farcm4$ to the east with radial velocity 
difference of $\Delta v_r \sim 300$\kms) may be in an early premerger 
phase with NGC\,6876. 

Important questions concerning the evolution of galaxies in the galaxy
group environment remain poorly understood. Among these are :
\begin{enumerate}
\item{How do galaxy interactions influence nuclear activity and black 
 hole growth in interacting galaxies in galaxy groups?}
\item{What are the patterns of star formation in interacting
  galaxies?}
\item{How do we observationally constrain the kinematics of
  galaxy-group interactions, such as the high velocity passage of  
  NGC\,6872 past NGC\,6876 in the Pavo group core, and how do  these 
  interactions influence the hydrodynamical state of both the group IGM and 
  the hot X-ray emitting gas in the galaxies?}
\end{enumerate}
In a series of three papers we address these questions for the
dominant interacting galaxies in the Pavo group. 
In a previous paper, we used {\it Spitzer} mid-infrared and {\it
  Chandra} X-ray images to analyze the role of the IC\,4970/NGC\,6872 
collision on the nuclear activity in the less massive galaxy, 
IC\,4970, arguing that it
hosts a highly-obscured active galactic nucleus 
(AGN) fueled by cold gas driven into the nucleus by the interaction  
(Machacek \etal 2008a). In the current paper, we analyze {\it Spitzer} 
mid-infrared observations of the dominant Pavo group galaxies, NGC\,6876 and
NGC\,6872, and {\it Chandra } observations of bright X-ray point
  sources found associated with the spiral galaxy 
NGC\,6872, to investigate the impact  of galaxy 
interactions on star formation in these systems. The detailed analysis of our 
{\it Chandra} observations of diffuse gas in and around these
  galaxies, that probes the impact of
  high velocity group-galaxy interactions on the hydrodynamical state
  and evolution of the Pavo group IGM, will be reported in a
  forthcoming paper (Machacek \etal 2008b, in preparation).

This paper is organized as follows:  
In  \S\ref{sec:obs} we briefly review the observations and our 
data reduction and processing procedures.
In \S\ref{sec:irac} we present our IRAC mid-infrared results, 
constructing flux density and nonstellar emission images of the 
NGC\,6876/NGC\,6877 and NGC\,6872/IC\,4970 galaxy pairs in the Pavo group,
along with IRAC band color maps of the more strongly interacting galaxy
pair, NGC\,6872/IC\,4970. 
In \S\ref{sec:xray} we compare the properties of the X-ray point
source population with the distribution of star forming regions 
identified in \S\ref{sec:irac}  
and with existing optical observations of young star clusters and
radio observations of the distribution of HI gas. We also comment on the 
relationship of extended  X-ray emission, observed with  
{\it Chandra} in NGC\,6872, with these star formation tracers in other 
wavebands. We briefly summarize our findings in \S\ref{sec:conclude}.

Unless otherwise indicated, uncertainties correspond
to  $90\%$ confidence levels on spectral parameters and $1\sigma$ 
uncertainties on X-ray counts and count rates. WCS coordinates are 
J2000. 
Adopting a $\Lambda$CDM cosmology with $H_0 = 73$ \kmsmpc and 
$\Omega_m = 0.238$ from the three year WMAP results (Spergel \etal
2006) and taking the redshift of the central group elliptical galaxy NGC\,6876 
($z=0.01338$) as representative of the redshift of the Pavo group, 
we find a luminosity distance of $55.5$\,Mpc for galaxies in the Pavo
group core. One arc second corresponds to a distance of 
$0.262$\,kpc. 


\section{Observations and Data Reduction}
\label{sec:obs}

Our mid-infrared results are based on the analysis of 
a $432$\,s observation of NGC\,6876 and NGC\,6872 ({\it Spitzer} PID 20440) 
taken on 2005 September 18 in the $3.6$, $4.5$, $5.8$, 
and $8.0\,\micron$ wavebands using the IRAC camera  
on board the {\it Spitzer} Space Telescope.
We used two IRAC pointings, one
with aim point centered on the dominant Pavo group 
elliptical galaxy NGC\,6876 ($\alpha=20^h18^m19.15^s$, 
$\delta=-70^\circ 51' 31\farcs7$, NED) 
and the other centered on the spiral galaxy NGC\,6872 
($\alpha=20^h16^m56.48^s$, $\delta=-70^\circ 46' 5\farcs7$, NED). 
 For a detailed  discussion of our 
mid-infrared data reduction and mosaicing procedures, please see 
Machacek \etal (2008a).  We obtained good coverage  with $>5$ Basic 
Calibrated Data (BCD) frames in all IRAC wavebands in and near 
the bright galaxies, NGC\,6876 and NGC\,6872. However,  
coverage of the region midway between NGC\,6876 and NGC\,6872 was 
sparse ($\gtrsim 2$ BCD frames). The BCD frames 
were super-boresight corrected and registered on the same sky grid, 
such that astrometric uncertainties were $\leq 0\farcs2$. 

Our X-ray data are from a $76$\,ks 
observation, taken in two segments on 2005 December 14-15 (OBSID 7248) and 
2005 December 16-17 (OBSID 7059) with the Advanced CCD Imaging
Spectrometer Imaging Array (ACIS-I; Garmire \etal 1992; Bautz \etal
1998) on board the {\it Chandra} X-ray Observatory.  
A detailed discussion of these {\it Chandra} X-ray observations, 
X-ray backgrounds and our data cleaning procedures, that resulted  in
useful exposure times of $31878$\,s and $38364$\,s for OBSID 7248
and OBSID 7059, respectively, are also reported in 
Machacek \etal (2008a). 

Point sources were identified in four X-ray energy bands, i.e. soft
($0.5-1$\,keV), medium ($1-2$\,keV), hard ($2-8$\,keV), and broad 
$0.5-8$\,keV energy bands, using both CIAO tool wavdetect with a
significance threshold of $10^{-6}$ and a
multiscale wavelet decomposition algorithm (wvdecomp) with $5\sigma$ 
detection threshold. We found no significant differences between the 
two methods. In addition to X-ray emission from the central regions of
the galaxies NGC\,6876, NGC\,6877, NGC\,6872, IC\,4970, and PGC\,06443
(shown in Fig. \ref{fig:pavodssgrp}), we find $64$ X-ray point sources to a
$0.5-8$\,keV flux limit of $\sim 1 \times 10^{-15}$\ergscm in 
a $236.5\,{\rm arcmin}^2$ field of view. We further checked our X-ray source
identification algorithms by comparing the number of sources we detect
to this flux limit with that expected from the Cosmic X-ray
Background. We detect $16$ X-ray sources in a $0.06^\circ$ radius
circle chosen on the ACIS-I field of view to exclude known galactic 
sources, in excellent agreement with the $18$ Cosmic X-ray
Background sources expected in this region to this flux limit 
(Brandt \etal 2001).

 Both segments of our {\it Chandra} observation 
(OBSID 7248 and OBSID 7059) placed the ACIS-I aimpoint near the 
spiral galaxy NGC\,6872 to achieve optimal angular resolution close to
the spiral galaxy, where effects of NGC\,6872's interactions are  
expected to be most prominent. However, this placed both the 
elliptical galaxies NGC\,6876 and NGC\,6877 $\sim 6\farcm8$ off axis, 
close to the edge of the I3 CCD, where the mirror point spread
function broadens and 
encircled energy radii become azimuthally asymmetric. Thus   
X-ray point sources, such as low-mass X-ray binaries that may be 
associated with NGC\,6876 or NGC\,6877 are difficult to distinguish from the
elliptical galaxies' hot diffuse X-ray gas halos. We found no additional 
X-ray point sources associated with either NGC\,6877 or NGC\,6868. 
Therefore, in this paper (\S\ref{sec:xray}), 
we discuss only the properties of the $12$ point sources likely 
associated with the spiral galaxy NGC\,6872. The properties of the 
active nucleus and near-nuclear point source in IC\,4970 
(NGC\,6872's companion galaxy) are the subject of Machacek \etal
(2008a). 

We compare our  {\it Spitzer} mid-infrared and {\it Chandra} X-ray
results with existing HI (Horellou \& Koribalski 2007),  
VLT B-band (Bastian \etal 2005), and H$\alpha$ (Mihos \etal 1993) 
observations from the literature.
  
\section{Pavo Galaxies Viewed in the Mid-Infrared}
\label{sec:irac}

In Figures \ref{fig:n6876mosaics} and \ref{fig:n6872mosaics} we
present background-subtracted mosaics of the dominant Pavo group
galaxy pairs NGC\,6876/NGC\,6877 and NGC\,6872/IC\,4970, respectively, 
in the four IRAC wavebands. The $3.6\,\micron$ and $4.5\,\micron$
wavebands (upper panels) are dominated by light from an old population
of stars. Emission in the $5.8$ and $8.0\,\micron$ bands is more complicated, 
containing possible contributions from PAHs and dust in young star-forming
regions, silicates ejected from the outer atmospheres of AGB
stars, or AGN accretion disks, as well as starlight. 
We separate the nonstellar from the stellar components at these 
wavelengths following the method of Pahre \etal (2004). 
Specifically, a model for the $5.8$ and $8.0\,\micron$ stellar emission 
is  constructed by averaging background-subtracted $3.6$ and
$4.5\,\micron$ flux density mosaics that have first been 
aperture corrected, using the IRAC extended (infinite) aperture
corrections provided by the {\it Spitzer} Science Center 
\footnote{ http://ssc.spitzer.caltech.edu/irac/calib/extcal/}, 
and then scaled to the appropriate waveband ($5.8\,\micron$ or
$8.0\,\micron$) 
using the 
approximate mid-infrared colors of an M0 III star 
($[3.6]-[4.5]=-0.15$, $[4.5]-[5.8] = 0.11 $, $[5.8]-[8.0] = 0.04$).  
This stellar model is then subtracted from the background-subtracted,
aperture-corrected, observed $5.8\,\micron$  and $8.0\,\micron$ mosaics to
produce the nonstellar emission maps, shown in the lower panels of   
Figures \ref{fig:n6876mosaics} and \ref{fig:n6872mosaics}. Below we 
discuss our mid-infrared results for elliptical galaxies NGC\,6876 and 
NGC\,6877 (\S\ref{sec:n6876iracflux}) and spiral galaxy NGC\,6872 
(\S\ref{sec:n6872iracflux}). The mid-infrared results for NGC\,6872's 
companion galaxy IC\,4970 were presented in Machacek \etal (2008a) and 
will not be repeated here.

\subsection{Dust in Elliptical Galaxies NGC\,6876 and NGC\,6877}
\label{sec:n6876iracflux}

\subsubsection{Mid-infrared Morphology}
\label{sec:n6876morp}

\begin{figure*}[t]
\begin{center}
\includegraphics[height=2.274in,width=3in]{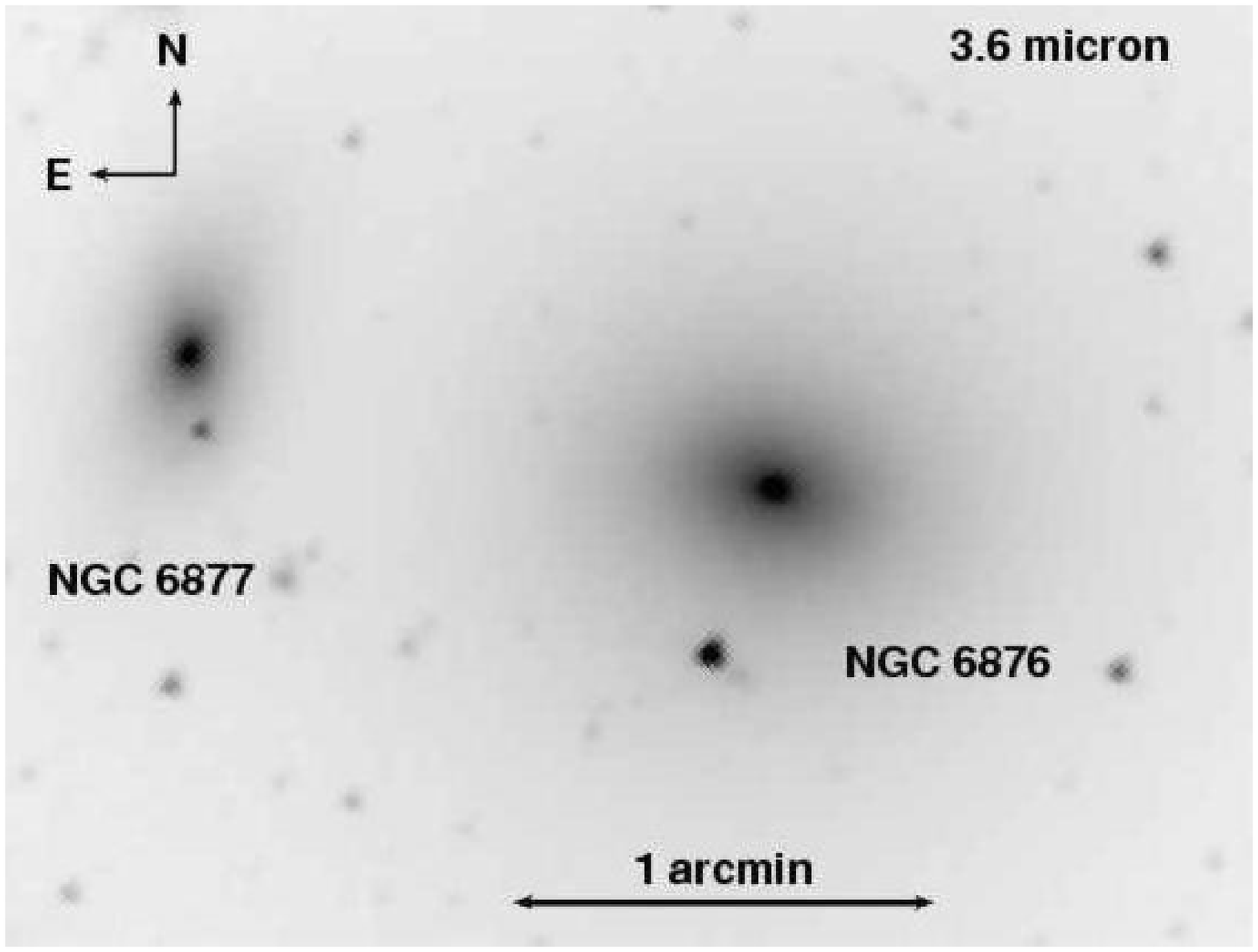}
\includegraphics[height=2.274in,width=3in]{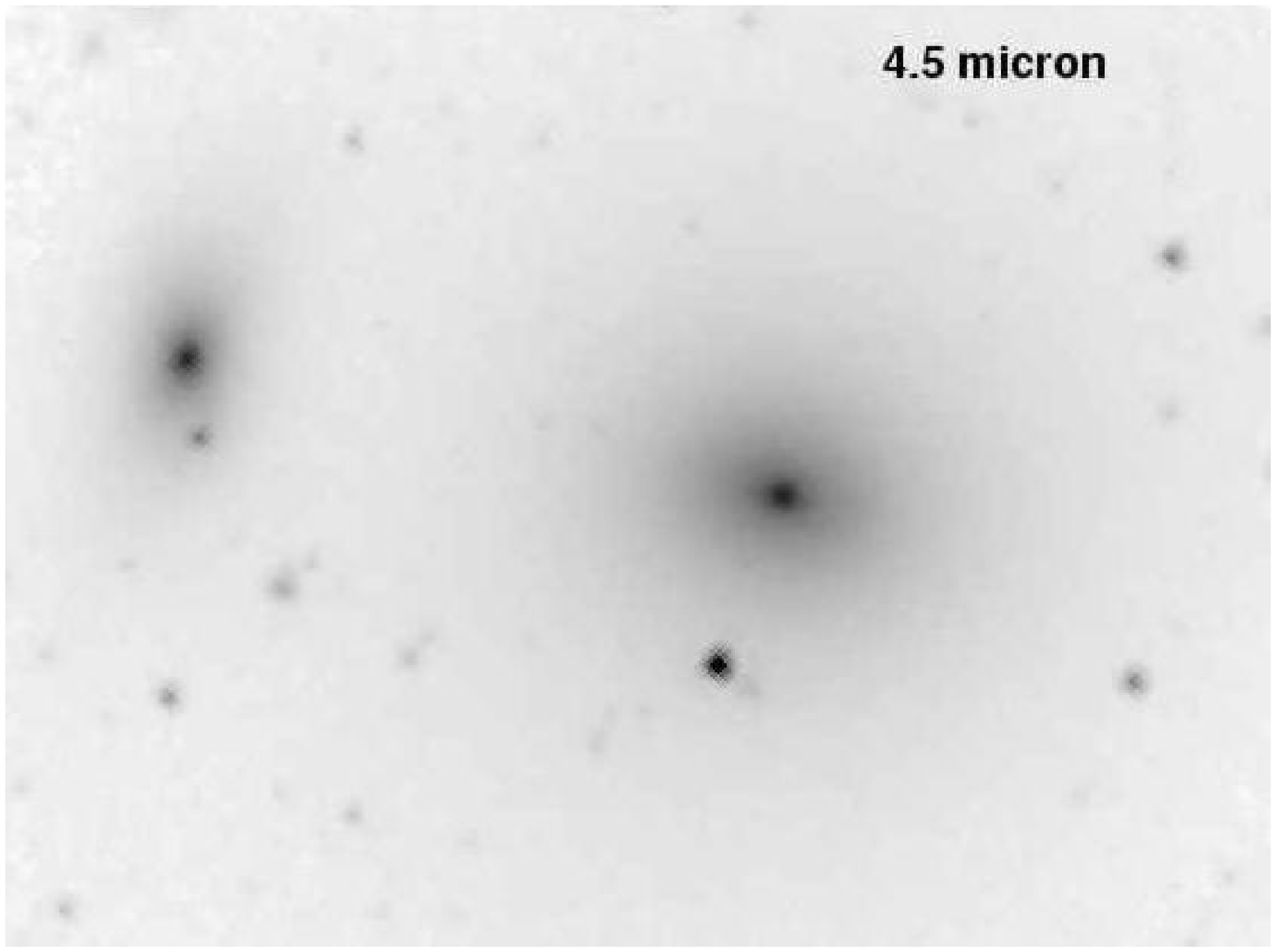}
\includegraphics[height=2.274in,width=3in]{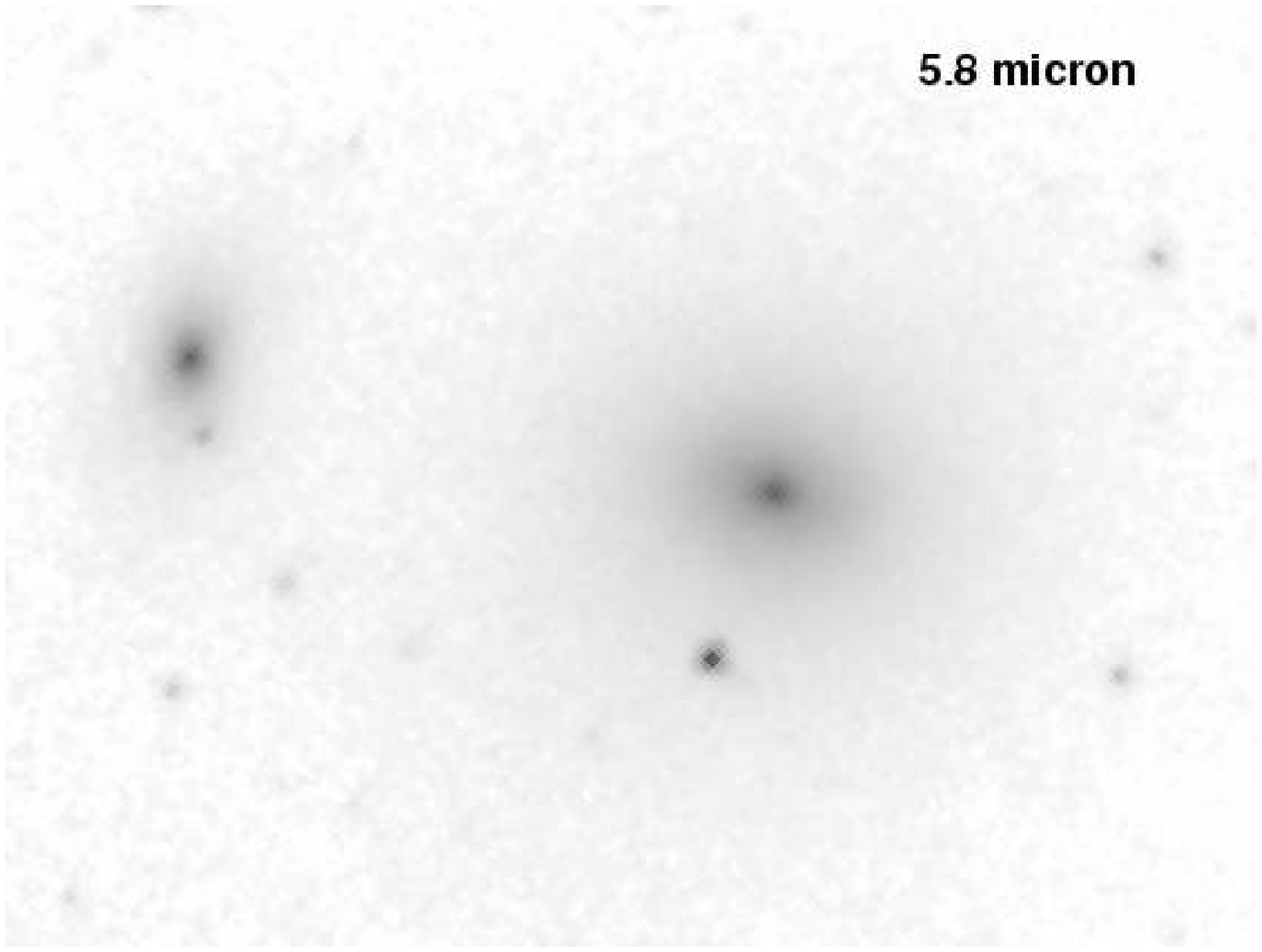}
\includegraphics[height=2.274in,width=3in]{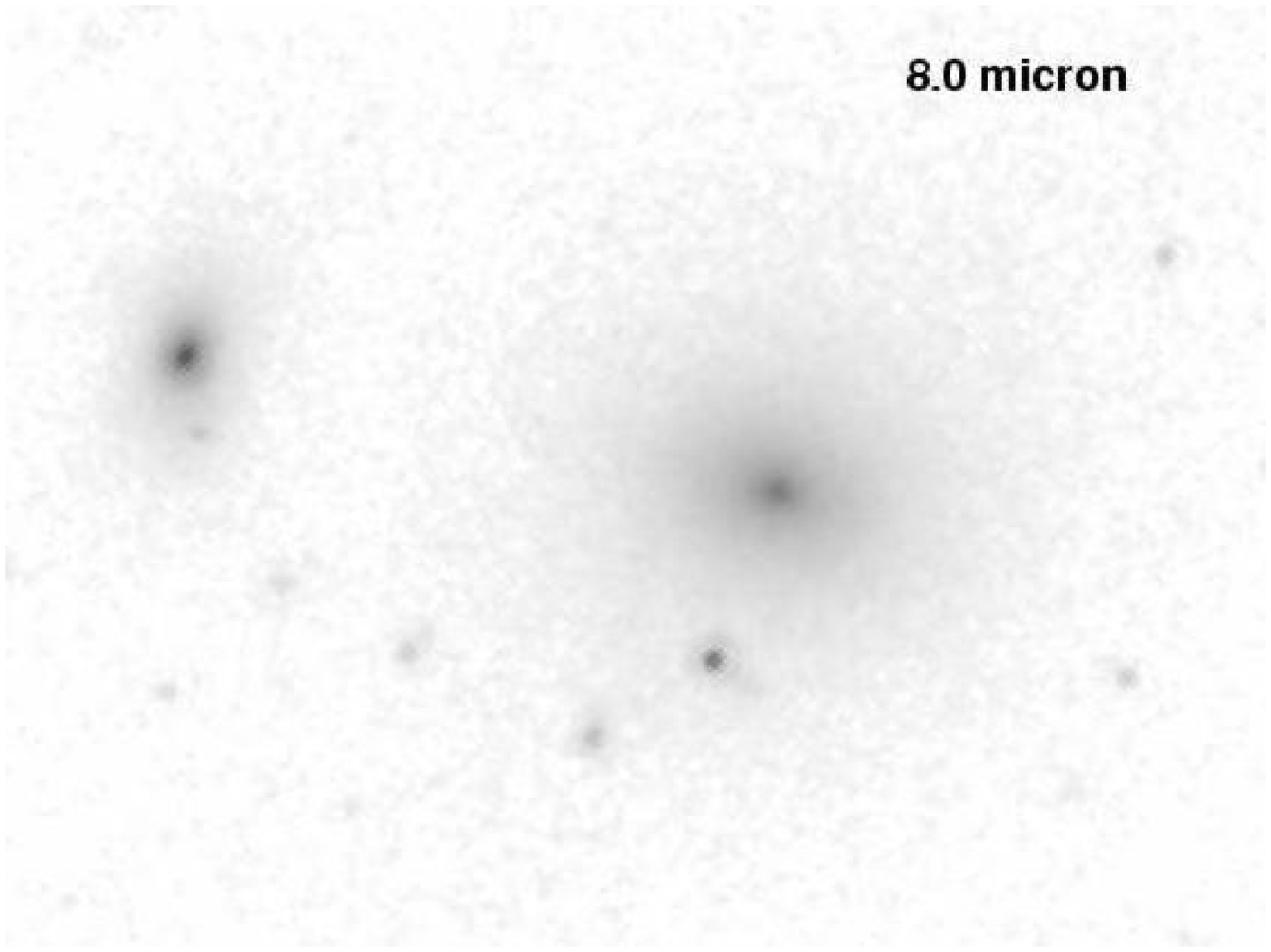}
\includegraphics[height=2.274in,width=3in]{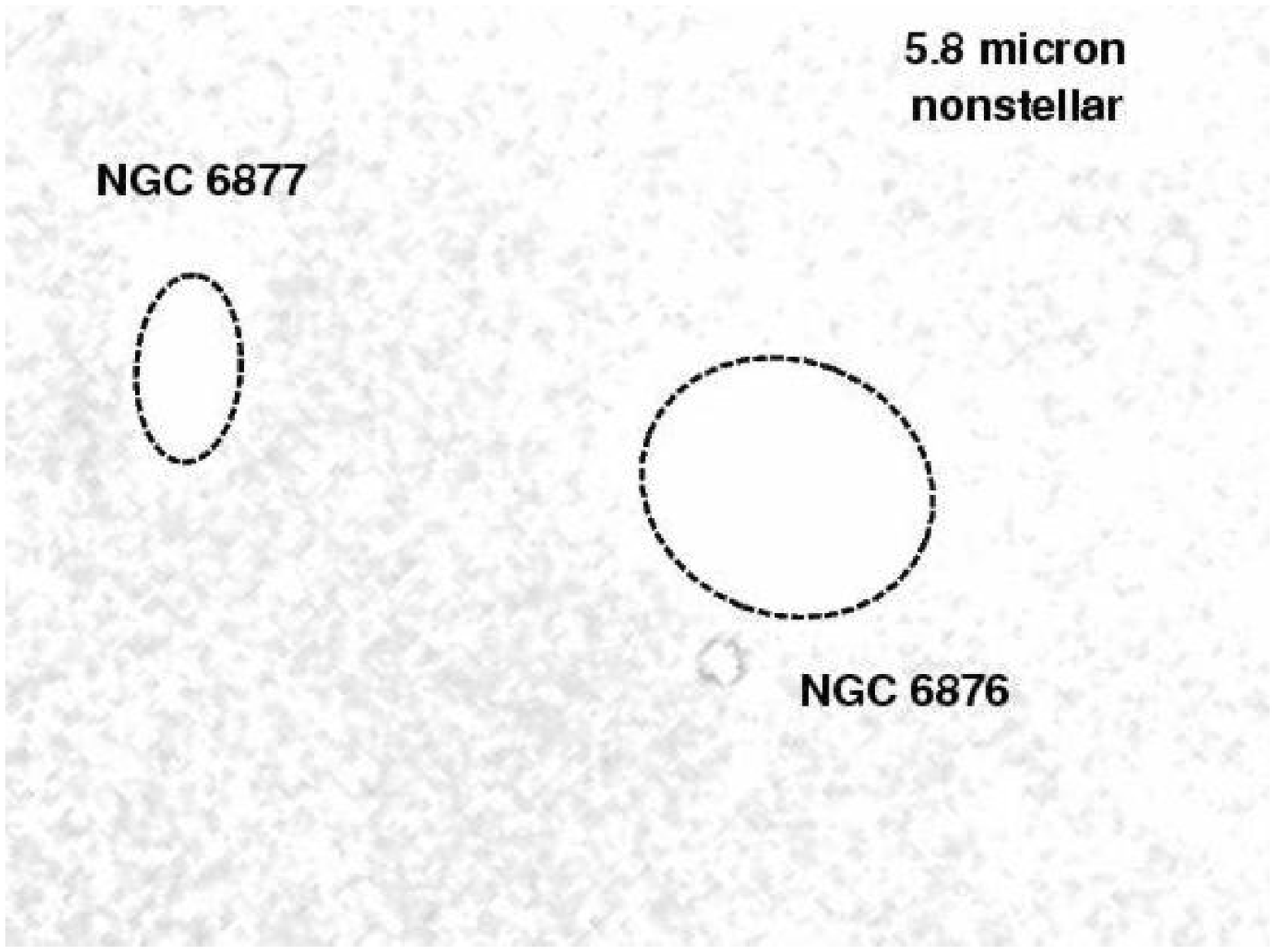}
\includegraphics[height=2.274in,width=3in]{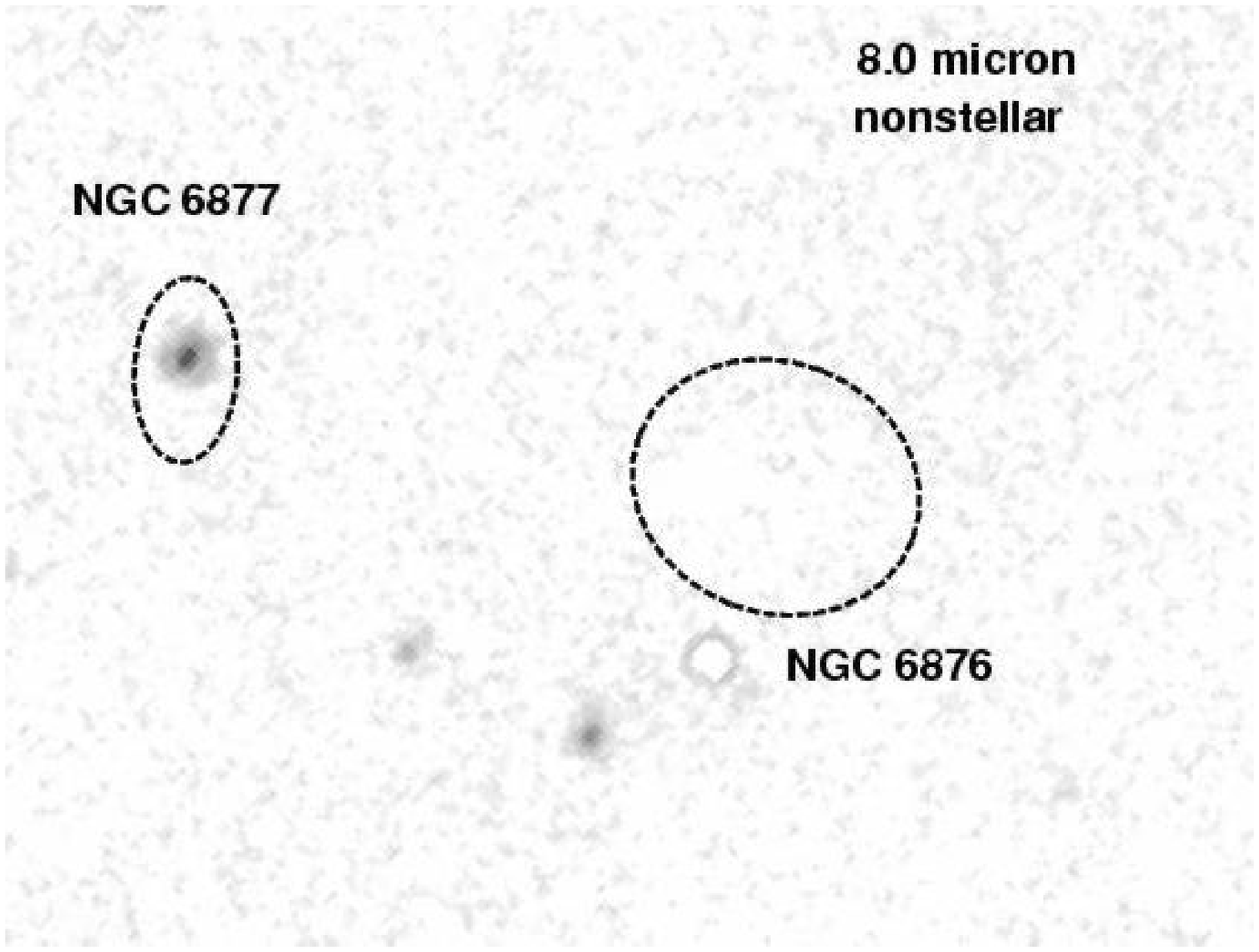}
\caption{ {\small Background subtracted, mosaiced Spitzer 
 images of the dominant elliptical galaxy NGC\,6876 and its companion
 galaxy NGC\,6877 in the $3.6\,\micron$ (upper left), 
 $4.5\,\micron$ (upper right), $5.8\,\micron$ (middle left) and 
 $8.0\,\micron$ (middle right) wavebands, respectively. 
 The greyscale is the same for upper and
 middle panels. Nonstellar emission maps of the dominant elliptical
 galaxy NGC\,6876 and its companion galaxy NGC\,6877 in the 
 $5.8\,\micron$ ( $8.0\,\micron$) wavebands are shown in the lower left
 (right) panels, respectively. Dashed lines denote the 
 $0.5$\mjysr contour level for NGC\,6876 and NGC\,6877 taken from the 
 $8.0\,\micron$ emission map (middle right). North is up and east is 
 to the left as shown in the upper left panel. }  
}
\label{fig:n6876mosaics}
\end{center}
\end{figure*}

Figure \ref{fig:n6876mosaics} shows that the surface brightness
distributions in all four IRAC wavebands are similar for 
NGC\,6876 and for NGC\,6877. 
For NGC\,6876, we find no evidence in the $5.8$ and $8.0\,\micron$ 
surface brightness maps for dust features that deviate from the
stellar light distribution in the elliptical galaxy. We present  
the nonstellar emission maps for NGC\,6876 and NGC\,6877 in the lower 
panels of Figure \ref{fig:n6876mosaics}. 
Nonstellar emission is seen in the 
central region of the nearby elliptical galaxy 
NGC\,6877, but the apparent lack of nonstellar emission in
NGC\,6876 is striking. However, this result 
should be interpreted with care, because IRAC aperture corrections for 
the $5.8$ and $8.0\,\micron$ wavebands are strongly dependent on the
size of the emission region and use of the limiting (large scale) 
aperture correction may underestimate the true flux by as much as
$25\%$ in these wavebands, if the emission is concentrated in smaller
regions.  Conservatively, 
the nonstellar emission maps for NGC\,6876 (lower panels of Fig.
\ref{fig:n6876mosaics}) suggest that dust emission, if present in the 
dominant group galaxy, is weak and likely follows the
galaxy's distribution of highly evolved stars.  

\subsubsection{Mid-infrared Photometry}
\label{sec:n6876sand}

We used fixed-aperture photometry in each IRAC waveband to 
investigate the possibility of dust emission in the elliptical
galaxies NGC\,6876 and NGC\,6877 more quantitatively. 
Our results and a complete description of the  source and background 
apertures used in the analysis are summarized in Table 
\ref{tab:n6876phot} for NGC\,6876 and Table \ref{tab:n6877phot} for its
companion elliptical galaxy NGC\,6877. The mid-infrared colors for
both galaxies are listed in Table \ref{tab:n6876colors}.
We define apertures for each galaxy as a whole (labeled `Full' in 
Tables \ref{tab:n6876phot} and \ref{tab:n6877phot}), based on the 
respective $0.5$\mjysr surface brightness contours 
in the $3.6\,\micron$ emission map (upper left panel of 
Fig. \ref{fig:n6876mosaics}). The background regions were taken to be 
concentric elliptical annuli immediately outside the source regions. 
Contaminating point sources, identified in the $3.6\,\micron$ map, were 
excised from both the source and background maps for each waveband.

The measured integrated flux 
densities for the `Full' regions were aperture corrected using 
correction factors for each
aperture's effective circular radius ($38\farcs5$ and $16\farcs6$ for 
NGC\,6876 and NGC\,6877, respectively) calculated 
from fits to the IRAC extended source calibration curves supplied by
the {\it Spitzer} Science 
Center\footnote{http://ssc.spitzer.caltech.edu/irac/calib/extcal/}.
Nonstellar emission, when present in
elliptical galaxies, is often centrally concentrated (Pahre \etal
2004), and may be masked by the more dominant stellar emission  
when averaged over the whole galaxy.  For example,  when averaged over the full
galaxy, neither the $5.8\,\micron$ nor the $8.0\,\micron$ nonstellar flux
densities, listed in Table \ref{tab:n6876phot} for NGC~6876 are greater than
the $\sim 10\%$ uncertainties. We thus use circular apertures
with radii of $6\farcs1$ and $3\farcs66$ (labeled `Center' in Tables 
\ref{tab:n6876phot} and \ref{tab:n6877phot}) to measure the mid-infrared flux 
densities in the central regions of NGC\,6876 and NGC\,6877, respectively. 
For such small apertures (i.e. radii $\lesssim 7''$), point
source aperture corrections appropriate for the chosen aperture size 
(IRAC Data Handbook 
version 3.0)\footnote{http://ssc.spitzer.caltech.edu/irac/dh/} 
were applied to the background-subtracted integrated flux densities to 
compute the `true' flux densities. Corresponding magnitudes for all cases were
calculated using the Vega relative zeropoint magnitudes from 
Reach \etal (2005). 

The mid-infrared colors for 
NGC\,6876, as a whole, are broadly consistent with the mid-infrared 
colors for an old (M0-K0) population of stars, as expected. 
However, our results for the  $5.8$ and $8.0\,\micron$ integrated 
flux densitites from the `Full' aperture 
are lower than those previously reported for NGC\,6876 in nearby wavebands 
by Ferrari \etal (2002) using observations from the 
Infrared Space Observatory (ISO). To check 
whether this discrepancy is due to differences in the chosen
apertures and/or the presence of point sources, excluded from our
analysis, that were unresolved in the ISO data, we measured the total 
integrated $5.8$ and $8.0\,\micron$ fluxes in a $108'' \times 108''$
square aperture centered on NGC\,6876 (based on the ISO field of view),
without excising point sources, and with aperture corrections 
appropriate for the $54''$ aperture effective radius. 
At $5.8\,\micron$ we find an integrated flux for this larger region of 
$102.5$\,mJy, in good agreement with the $125$\,mJy observed 
at $6.75\,\micron$ by Ferrari \etal, given the $30\%$ 
calibration uncertainties in the ISO data. However, for 
$8.0\,\micron$ we find an 
integrated flux in this larger aperture of $50.5$\,mJy, still a factor 
$\sim 3$ below the $192$\,mJy measured by Ferrari and collaborators in the
$9.63\,\micron$ waveband. This may be further
evidence, as suggested by Ferrari \etal (2002), that the excess
emission at $9.6\,\micron$ is due to warm silicate grains from the
atmospheres of old AGB stars, and not PAH emission from young star 
forming regions. Such an interpretation would be consistent with
results for elliptical galaxies in richer environments, such as the 
Virgo cluster, where $\sim 82\%$ of the $17$ early type galaxies 
studied by Bressan \etal (2006; 2008) with ISO mid-infrared
spectroscopy showed silicate emission.

 While the $[3.6]-[4.5]$ and $[4.5]-[5.8]$ colors for the central
 region of NGC\,6876 (labeled `Center' in Table \ref{tab:n6876colors})
 are again broadly consistent with an old population of stars, 
the $[5.8]-[8.0]$ color is redder than expected. However, after 
subtracting our stellar model to isolate the nonstellar contribution
to these central flux densities,  
we do not find a statistically significant
($> 10\%$) contribution to the flux from nonstellar
emission in the $5.8\,\micron$ band, and find only a weak ($16\%$) 
contribution from nonstellar emission in the $8.0\,\micron$
band. This is again consistent with the interpretation that the excess 
$8.0\,\micron$ flux is due to leakage from the broad $9.7\,\micron$ 
emission feature expected for warm silicate dust grains into the
$8.0\,\micron$ bandpass. In normal star-forming galaxies the star
formation rate is correlated with the $8.0\,\micron$ emission from 
warm dust by the expression (Wu \etal 2005)
\begin{equation}
{\rm SFR}(\sfr) = \frac{\nu L_{\nu}(8\,\micron,{\rm nonstellar})}{1.57
  \times 10^9\,L_\odot}\,\,.
\label{eq:sfr}
\end{equation}
Using Equation \ref{eq:sfr}, we find an upper limit on star formation 
in NGC\,6876 of $\lesssim 0.02\sfr$.

Figure \ref{fig:n6876mosaics} does show nonstellar emission in the 
central region of NGC\,6877, the companion galaxy to NGC\,6876. In
particular, the $[3.6] - [8.0]$  color for NGC\,6877 (expected to be 
close to zero for old stars) is $0.27$,  
too red to be caused by any stellar population (Pahre \etal 2004). 
After subtracting the contribution from our stellar model, we find 
that nonstellar emission contributes $47\%$ ($14\%$) of the
$8.0\,\micron$ ($5.8\,\micron$) flux densities, respectively, in the
central region of NGC\,6877. The nonstellar $[5.8] - [8.0]$ color of
$1.86$ in the central region is similar to the $[5.8] - [8.0]$
nonstellar colors ($\sim 1.42-2$) found in star-forming regions in M81
(Willner \etal 2004) and in dusty E/S0 galaxies (Pahre \etal 2004)
and is consistent with that expected from 
PAH molecules and dust in star-forming regions (Li \& Draine 2001). However, 
the observed (stellar plus nonstellar) $[3.6]-[4.5]$ and $[5.8]-[8.0]$ 
colors for the central region of NGC\,6877 place the
galaxy close to normal S0/Sa galaxies in the color-color diagram of 
Stern \etal (2005), such as for NGC\,4429 in which  star formation 
rates are low. Using the $8\,\micron$ nonstellar flux density for 
NGC\,6877 in Equation \ref{eq:sfr}, we estimate a star 
formation rate of $0.04\sfr$. 
Thus, as in NGC\,4429, star formation in the central
region of NGC\,6877, while likely present,  is weak.
 
\subsection{Star Formation in Spiral Galaxy NGC\,6872}
\label{sec:n6872iracflux}

\begin{figure*}[t]
\begin{center}
\includegraphics[height=2.274in,width=3in]{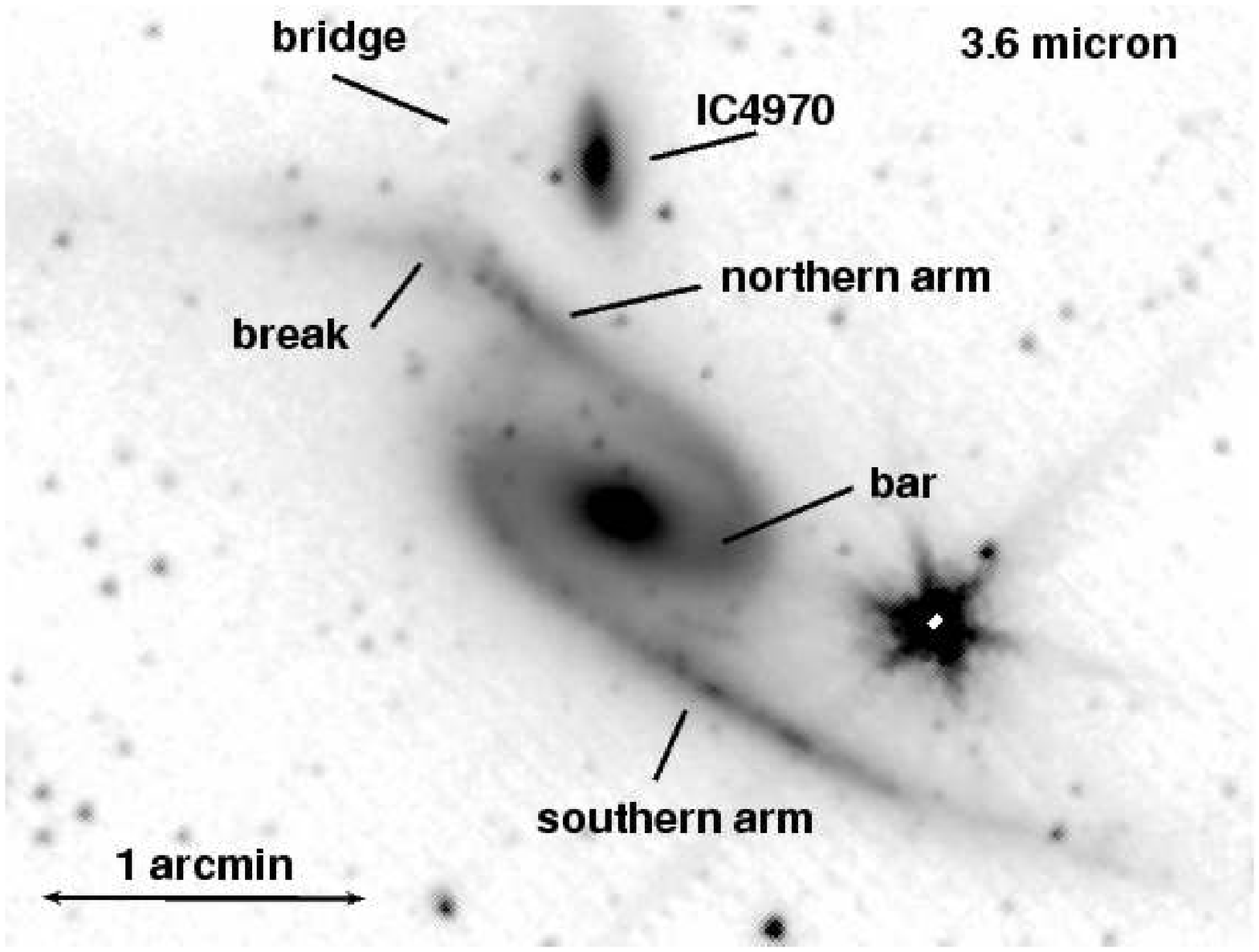}
\includegraphics[height=2.274in,width=3in]{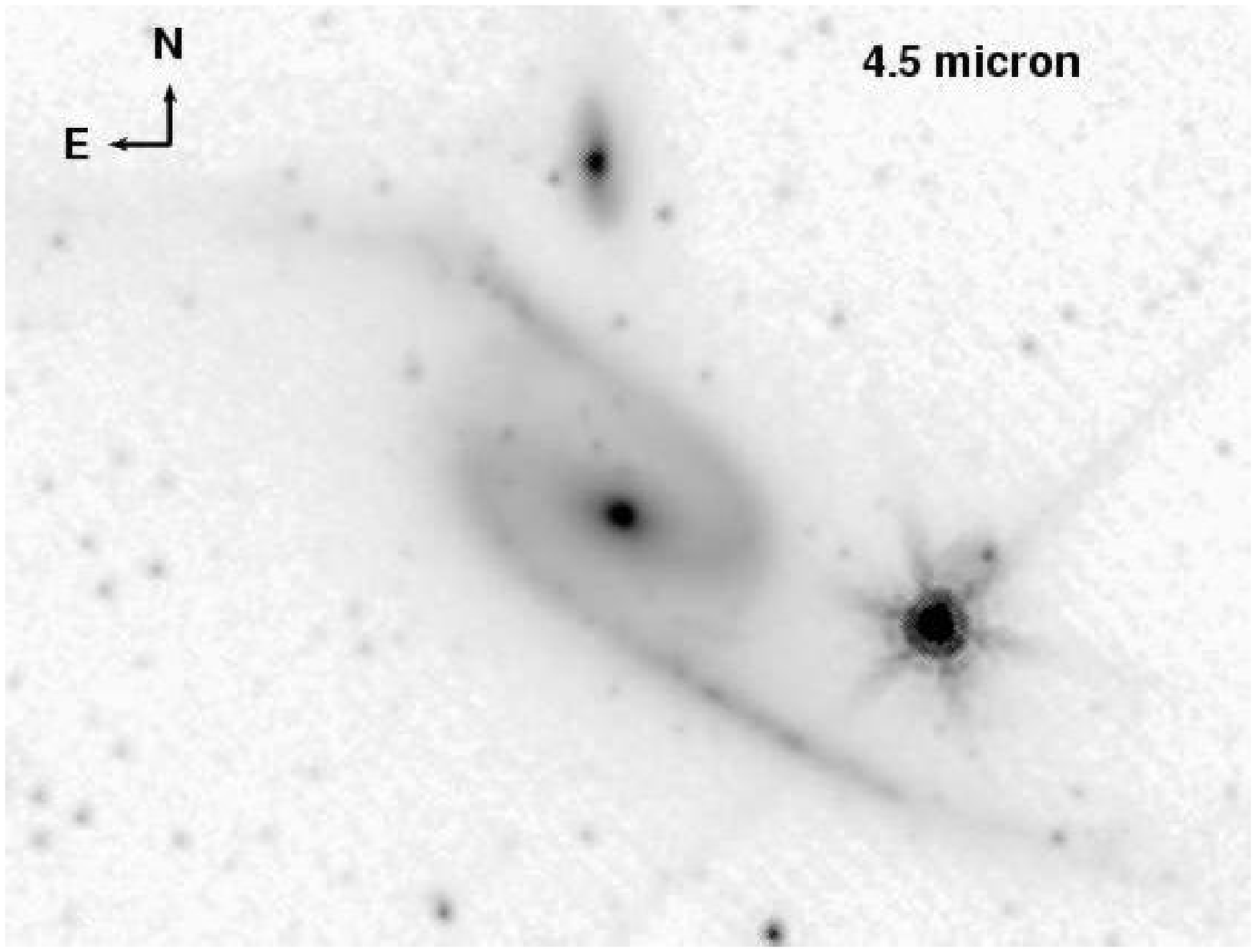}
\includegraphics[height=2.274in,width=3in]{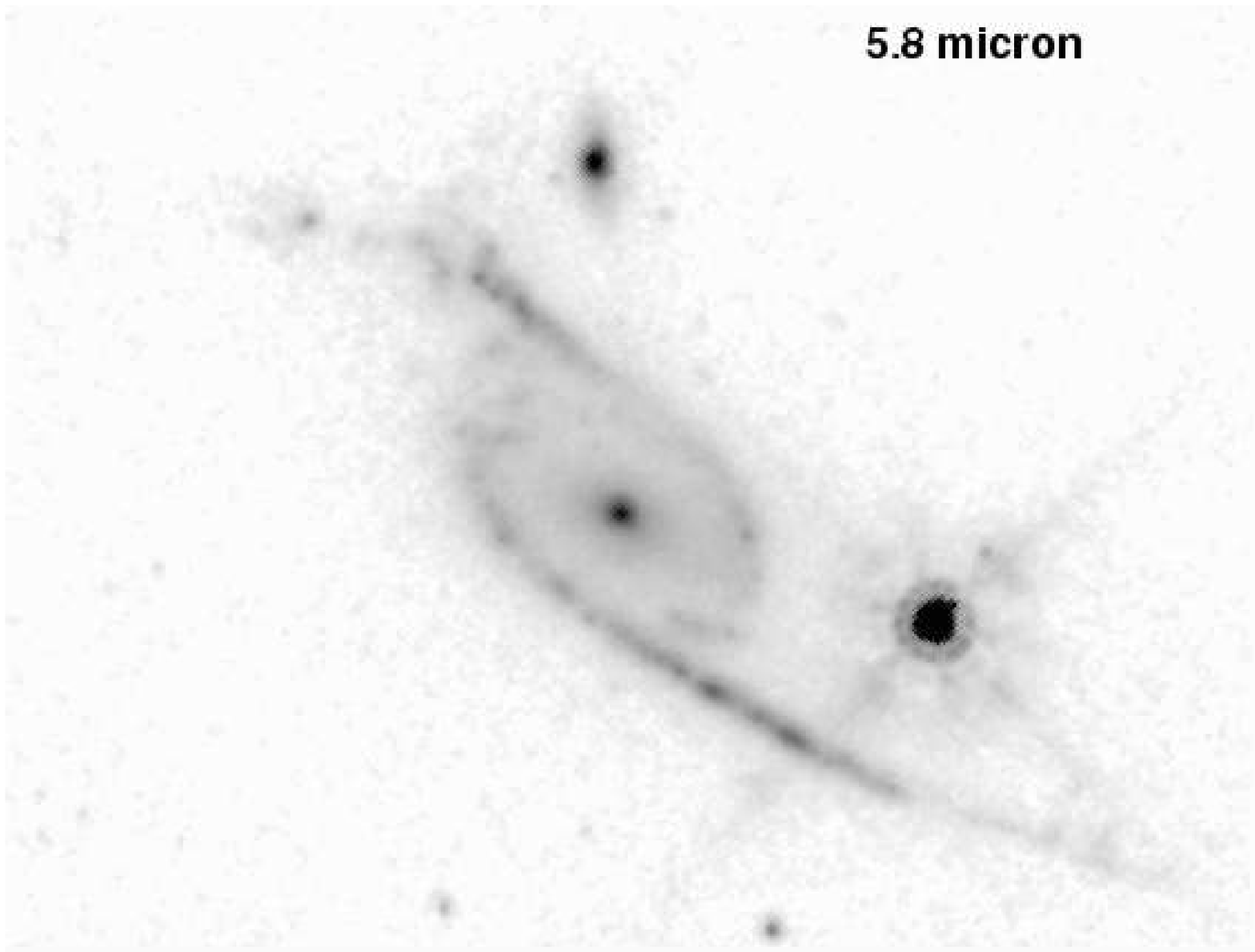}
\includegraphics[height=2.274in,width=3in]{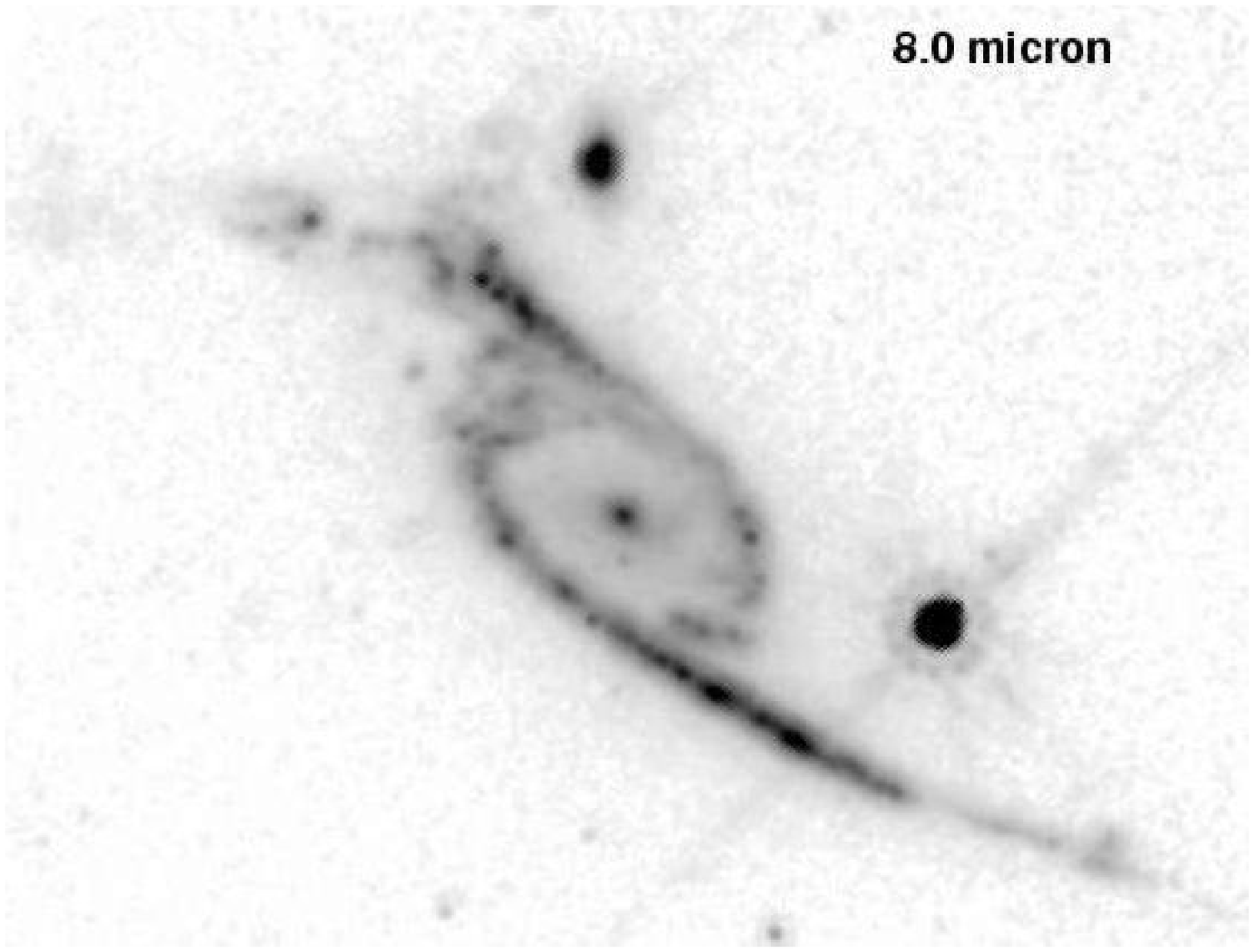}
\includegraphics[height=2.274in,width=3in]{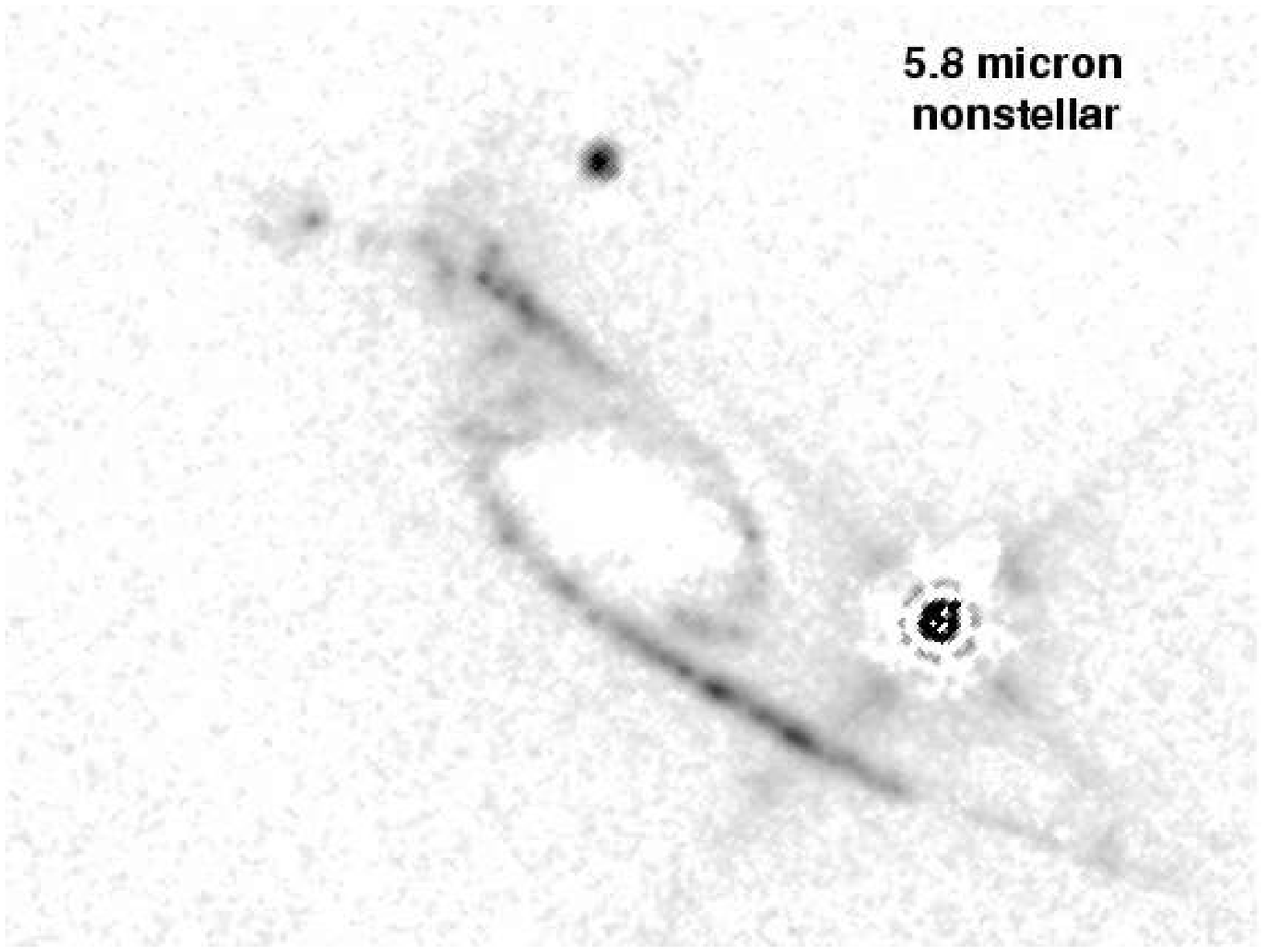}
\includegraphics[height=2.274in,width=3in]{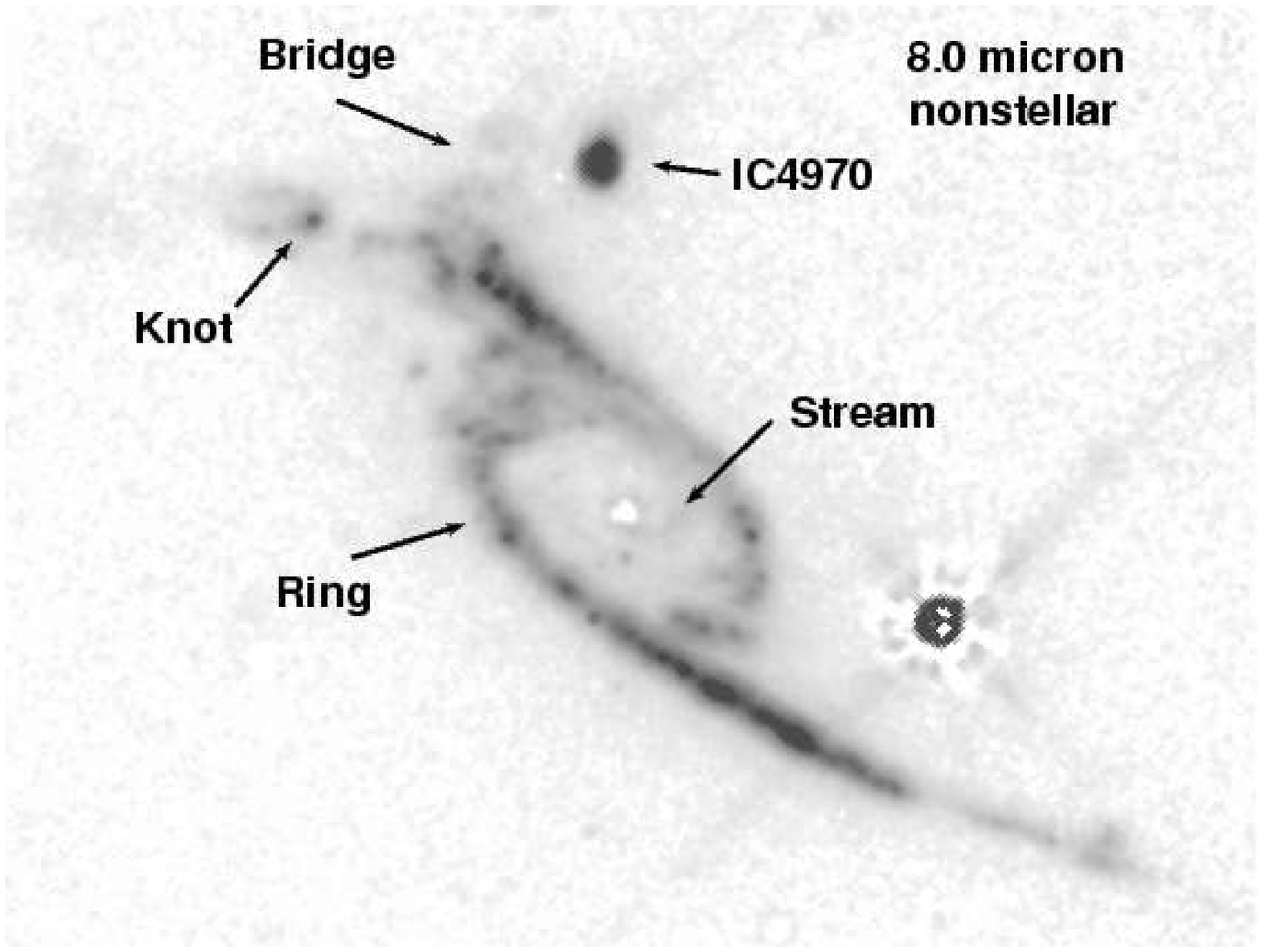}
\caption{ {\small Background subtracted, mosaiced {\it Spitzer}  
 images of the large spiral galaxy NGC\,6872 and its interacting companion
 spheroidal galaxy IC\,4970 in the $3.6\,\micron$ (upper left), 
 $4.5\,\micron$ (upper right), $5.8\,\micron$ (middle left) and 
$8.0\,\micron$ (middle right) wavebands.
 The greyscale 
is the same for the upper and middle rows. Nonstellar emission maps 
of the spiral galaxy NGC\,6872 and its companion galaxy IC\,4970 in the 
 $5.8\,\micron$ ($8.0\,\micron$) wavebands are shown in the lower left
 (right) panels, respectively. North is up and east is to the left 
  as shown in the upper right panel. } 
}
\label{fig:n6872mosaics}
\end{center}
\end{figure*}

From Figure \ref{fig:n6872mosaics} we see that the surface brightness
distribution in the $3.6\,\micron$ and $4.5\,\micron$ wavebands, that
trace the distribution of starlight in 
NGC\,6872, are very different from the $5.8\,\micron$ and
$8.0\,\micron$ surface brightness maps of the large spiral galaxy. 
The $3.6\,\micron$ and 
$4.5\,\micron$ maps (upper panels) show a bright nucleus, clear
bar and two tidally distorted spiral arms, similar to the Bj-band
image shown in Figure \ref{fig:pavodssgrp}. In contrast,  
the $8.0\,\micron$ and $5.8\,\micron$ emission  in the spiral galaxy
NGC\,6872 (middle panels) is concentrated in an outer ring
and along both tidal tails. 
As shown in the lower panels of Figure \ref{fig:n6872mosaics}, these 
longer wavelength bands are dominated by nonstellar emission.  
This nonstellar emission tends to be clumpy, with clump sizes ranging 
from those unresolved by IRAC, i.e  $\lesssim 0.5$\,kpc in diameter, to
complexes $\gtrsim 1$\,kpc across. 
Along the northern tidal arm, bright nonstellar emission 
is found
concentrated between the ring and the break in the arm, along the 
bridge connecting the break to the companion galaxy IC\,4970, in a 
bright two-tailed knot east of the break, 
and in faint patches
extending out $> 2'$ to the full extent of the northern tidally distorted 
stellar tail (see also Fig. \ref{fig:nonstellarcompare}).
Nonstellar emission is found in the southern tidal arm in 
bright $1-2$\,kpc long beads west of the ring and again in fainter 
patches out to the end of the southern tidal tail. 
Since the same emission features that are seen in the nonstellar 
$8.0\,\micron$ map of NGC\,6872  also appear in the corresponding 
$5.8\,\micron$ map (albeit with lower signal to noise), the observed
flux is likely the result of PAH and dust emission from young, 
dust-enshrouded  star-forming regions in the spiral galaxy. 

\subsubsection{Mid-infrared Colors of the Spiral Arms and Tidal Features}
\label{sec:mircolors}

\begin{figure}[t]
\begin{center}
\includegraphics[height=1.97in,width=3.0in]{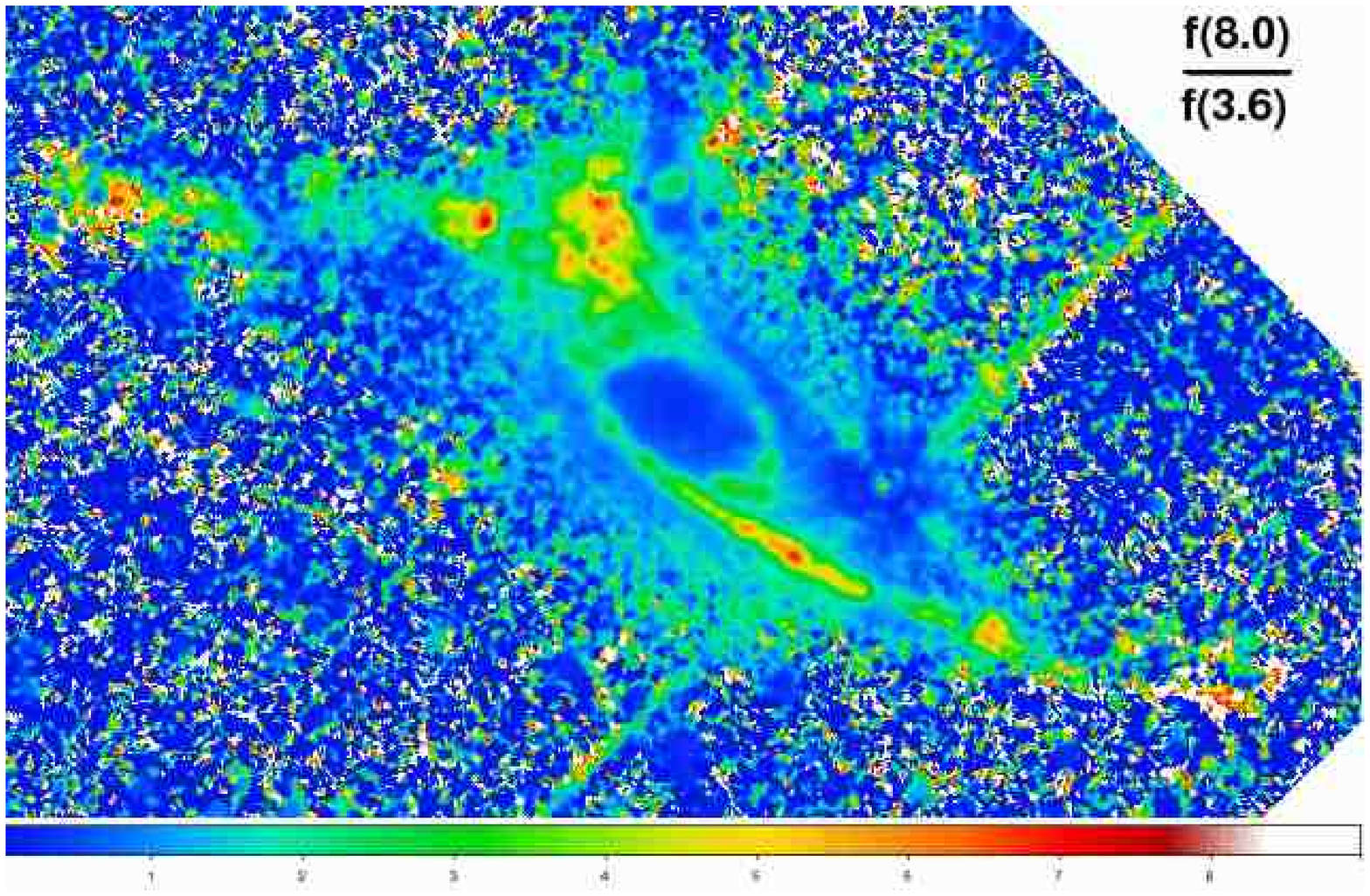}
\includegraphics[height=1.97in,width=3.0in]{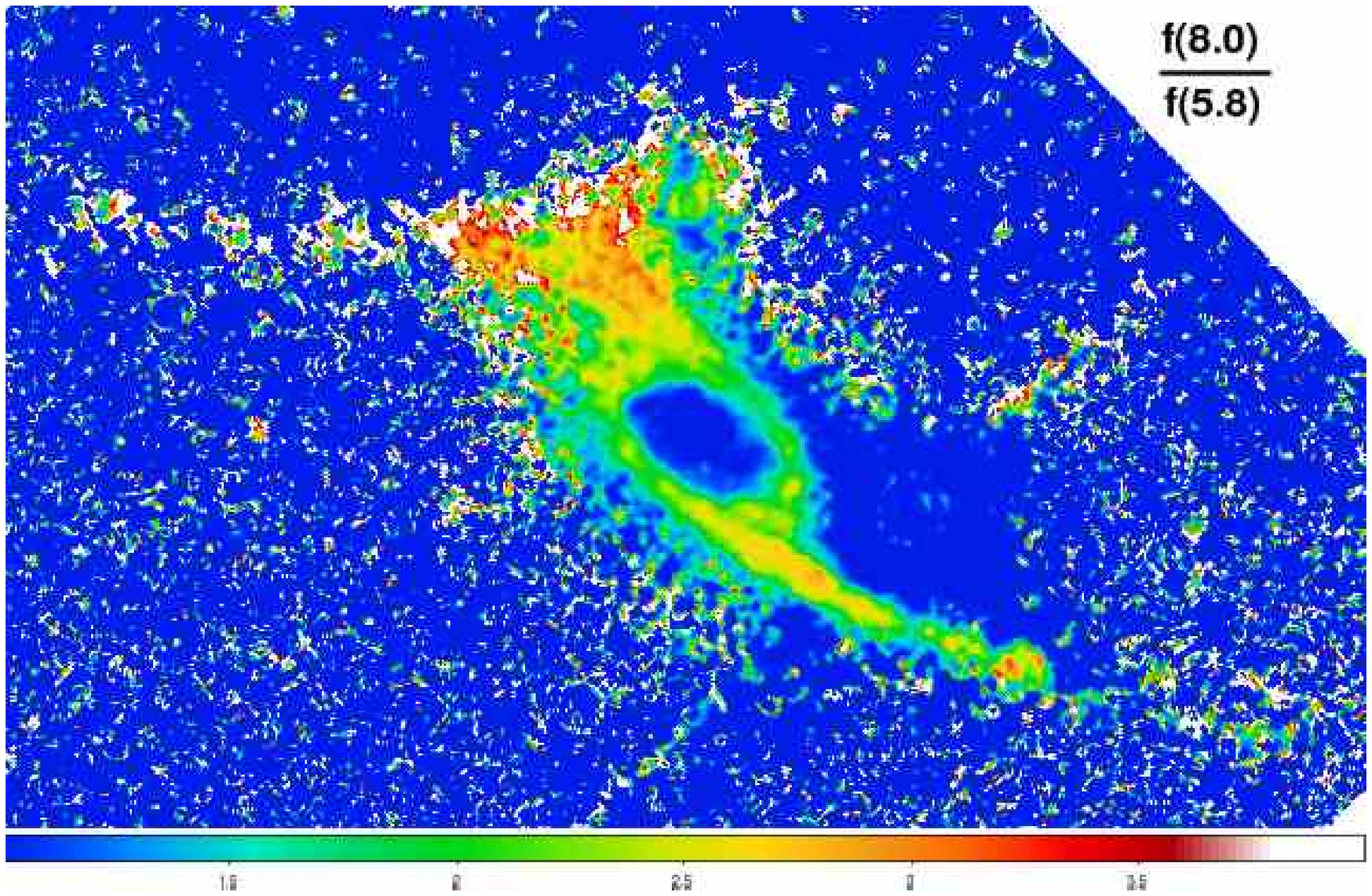}
\includegraphics[height=1.97in,width=3.0in]{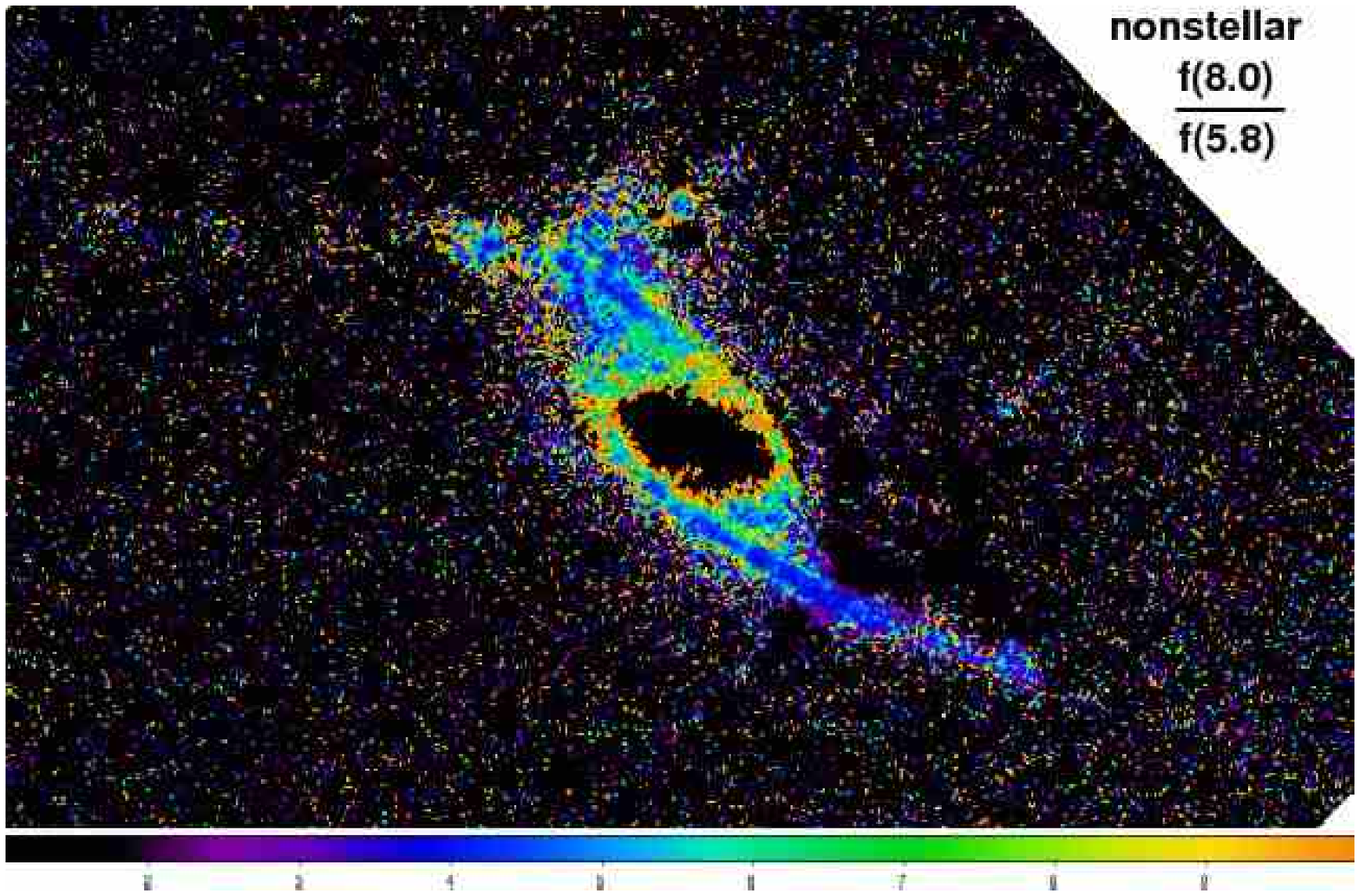}
\caption{{\small Mid-infrared  flux density ratio maps of NGC\,6872. 
The maps in each waveband were
cross-convolved with Gaussian representations of the point spread
functions before division to mitigate light scattering
effects. 
 North is up and east is to the left.
 Mid-infrared color magnitude scales for the maps are 
(upper)  
 $[8.0]-[3.6] = 0.30$ (blue), $2.27$ (cyan), 
 $2.80$ (green), $3.24$ (yellow), $3.55$ (orange), $3.72$\,mag. (red); 
(middle) $[8.0]-[5.8] = 0.35$ (blue), $1.11$ (cyan), 
 $1.42$ (green), $1.70$ (yellow), $1.86$ (orange), $1.96$\,mag. (red);
(lower) Blue (green) correspond 
 $([5.8]-[8.0])_{\rm nonstellar} = 2.1$ (blue), $2.7$\,mag (green). } 
 }
\label{fig:colormaps}
\end{center}
\end{figure}

We test the hypothesis that the observed
nonstellar emission is from PAH molecules and warm dust by using 
mid-infrared color maps to study the properties of the mid-infrared 
emission throughout the spiral galaxy and its tidal features. We show 
three representative color maps in Figure \ref{fig:colormaps}.
The background-subtracted mosaics in each waveband
were cross-convolved with a Gaussian representation of the IRAC point
spread function in the other waveband before division to
mitigate light scattering effects. We estimated the uncertainties in
the color maps  by comparing the colors of NGC\,6872's nucleus
obtained directly from the color maps to those calculated from the
flux densities measured by fixed aperture photometry in 
\S\ref{sec:nuclearphot} and find that the flux ratios (colors) 
obtained from Figure \ref{fig:colormaps} differ from
those obtained from fixed aperture photometry by $\lesssim 5\%$ 
($0.06$\,mag), comparable to  the calibration uncertainties ($\sim
10\%$) for these wavebands. 

In the central region and along the bar of the spiral galaxy, the 
$3.6\,\micron$ to $4.5\,\micron$ flux density ratio is 
 $S_{3.6}/S_{4.5} \sim 1.7$ to $1.85$, corresponding to  
$[3.6]-[4.5]$ colors of$-0.09$ to $-0.18$, respectively. These colors 
are  consistent with that expected for an old population of stars. 
The bright nonstellar emission regions, that are
particularly visible in the southern arm of NGC\,6872 in the lower
right panel of Figure \ref{fig:n6872mosaics} 
and in Figure \ref{fig:colormaps}, 
have $S_{3.6}/S_{4.5} \sim 1.4$, lower 
than elsewhere in the galaxy. The corresponding $[3.6]-[4.5]$
color ($0.12$) is in agreement  with the colors of star-forming 
clumps in the spiral and tidal bridge of NGC\,2536 (Hancock \etal 2007).
Depressed $S_{3.6}/S_{4.5}$ flux ratios were also found by Elmegreen \etal
(2006) in star-forming clumps in the NGC\,2207/IC\,2163 interacting
galaxy pair. They suggested that the low $S_{3.6}/S_{4.5}$ ratio may be due to 
an excess of H Br$\alpha$ emission at $4.5\,\micron$ caused by the 
ionization of
gas by hot, bright stars inside the star-forming clumps, such as 
found  in RCW 49, one of the most luminous HII regions in the Milky
Way (Whitney \etal 2004). 

The $S_{5.8}/S_{4.5}$ flux ratio ($[4.5]-[5.8]$ color) for NGC\,6872 
provides a measure of the relative number of young to old stars 
in the region (Smith \etal 2005a). This flux ratio (color) 
increases (reddens) from $\sim 0.7-0.8$ ($\sim 0.1-0.2$ mag) in the 
central region of the galaxy, consistent with an old population of stars,  
 to $\sim 1.4$ ($\sim 0.84$ mag) in the outer ring, indicating a younger 
stellar population and/or more interstellar dust. The bright
nonstellar emission regions are redder still, 
with $S_{5.8}/S_{4.5} > 3$ ($[4.5]-[5.8] > 1.7$) for clumps 
embedded in the tidal arms and bridge, and $S_{5.8}/S_{4.5} \sim 3.9$ 
($[4.5]-[5.8] \sim 1.9$) in the bright knot east of the break in 
the northern arm. These colors are consistent with those found in
star-forming regions in other interacting galaxies (Smith \etal 2005;
Hancock \etal 2007; Elmegreen \etal 2006).

The $S_{8.0}/S_{3.6}$ flux ratio ($[3.6]-[8.0]$ color), shown in the 
upper panel of Figure \ref{fig:colormaps}, provides a
measure of the star formation rate per unit stellar mass throughout
the galaxy. 
 The $S_{8.0}/S_{3.6}$ flux ratios in the southern 
tidal arm and bridge ($\sim 6 - 7$) and in the bright nonstellar
emission knot in the northern tidal arm east of the bridge ($7.6$) are
a factor $\sim 20$ higher than that found at the  center of 
NGC\,6872 (see Table \ref{tab:n6872col}). 
These flux density ratios  correspond to $[3.6]-[8.0]$ colors of 
$ \sim 3.6$, and $4.3$, respectively, approaching the
theoretical prediction of $4.95$ for interstellar dust (Li \& Draine
2001), and lie at the upper end of the 
distribution for $[3.6]-[8.0]$ colors in tidal features and M51-like
disks observed in the SSB\&T sample (Smith \etal 2007).

The $S_{8.0}/S_{5.8}$ flux ratio ($[5.8]-[8.0]$ color), 
shown in the middle panel of Figure \ref{fig:colormaps}, probes
the  properties of the emitting dust. In the southern tidal arm,  
in the northern tidal arm near the break and in the bridge, where there is
good signal-to-noise, we find $S_{8.0}/S_{5.8} \sim 2.6 - 3.25 $ 
($[5.8]-[8.0] \sim 1.8$) similar to that observed in other interacting
systems (see, e.g., the Antennae galaxies NGC\,4038/NGC\,4039, 
Wang \etal 2004; 
NGC\,2536, Hancock \etal 2007; NGC\,2207/IC\,2163, Elmegreen \etal 2006;  
Arp 107, Smith \etal 2005a; and M51-like disks and tidal
features in the SSB\&T sample, Smith \etal 2007). Once the stellar 
component is removed (see the lower panel of
Fig. \ref{fig:colormaps}), the $[5.8]-[8.0]$ color of the nonstellar
emission in these regions is $\sim 2.1$, in excellent agreement with 
the theoretical prediction of $2.06$ for pure PAH emission (Li \& Draine 2001).

Taken together these maps show a consistent picture of PAH and warm
dust emission from interaction induced star formation occuring 
in clumps or knots extending 
out to $\sim 50$\,kpc from the nucleus to the northeast and southwest 
along the tidal tails, but avoiding the nucleus and central $5$\,kpc 
of NGC\,6872. Regions of intense star formation, with  
mid-infrared colors similar to those observed in other 
interacting galaxies, are found both in the northern arm and bridge, 
near the ongoing interaction with IC\,4970, and in the southern tidal
arm, on the side of NGC\,6872 farthest from the present interaction. 
These features are strongly correlated with H$\alpha$ emission
in NGC\,6872 and, like the H$\alpha$ emission, are bright where the 
stellar velocity dispersion is high, further evidence that star
formation in this system is collisionally induced (Mihos 1993).  
We see no strong differences in the mid-infrared colors of the star-forming 
regions in NGC\,6872 as a function of their position relative to the 
companion galaxy IC\,4970. 

\subsubsection{No Star Formation in  NGC\,6872's Nuclear Region?}
\label{sec:nuclearphot}

Although strong nonstellar emission is found in the nuclear region of the 
companion galaxy, IC\,4970, 
consistent with the presence of a highly obscured AGN (Machacek \etal 2008a),
nonstellar emission from the nucleus of the primary galaxy in the
collision, NGC\,6872, is weak. 
From the color maps in Figure \ref{fig:colormaps} we see that,   
in the central region of NGC\,6872, the 
$S_{8.0}/S_{5.8}$ flux ratio ($\sim 0.7$) and corresponding 
$[5.8]-[8.0]$ color ($ \sim 0.25$) are modestly 
higher than expected for a population of old stars. 
As in \S\ref{sec:n6876iracflux} we use fixed aperture
point source photometry of NGC\,6872's nucleus using a $3\farcs66$ 
circular source aperture centered at the peak of the $3.6\,\micron$ emission 
($20^h16^m56.46^s$, $-70^\circ46'4\farcs80$, J2000) with a concentric 
background annulus with inner (outer) radii of $3\farcs66$
($8\farcs54$), respectively, to measure  
the fluxes and colors of NGC\,6872's nucleus and isolate any possible 
contribution from dust. Our results for the fluxes are presented in 
Table  \ref{tab:n6872phot}. 
In Table \ref{tab:n6872col} we list 
our photometric results for the mid-infrared flux ratios and colors 
of NGC\,6872's  nuclear region and compare them to the flux ratios and colors
obtained from the color maps in Figure  \ref{fig:colormaps}. 
We find excellent agreement between the colors determined by the two methods. 
 The mid-infrared colors fall in the region of the
color-color diagrams of Lacy \etal (2004) and Stern \etal (2005)
populated by normal (nonactive) galaxies that are dominated by starlight. 
The $[3.6]-[4.5]$ and $[4.5]-[5.8]$ colors for the 
nuclear region of NGC\,6872 are consistent with those expected for an 
old population of stars. The $[5.8]-[8.0]$ color ($ \sim 0.25$) is 
modestly higher suggesting the presence of dust. 
Subtracting a stellar model (as in \S\ref{sec:n6876iracflux}), we find  
that the remaining nonstellar component contributes $\sim 3\%$ 
($0.15$\,mJy) of the total emission in the $5.8\,\micron$ band, i.e. 
consistent with zero within calibration uncertainties, and $\sim 22\%$ 
($0.79$\,mJy) of the total emission in the $8\,\micron$ band. 
If, instead,  we model the
distribution of stars using the $3.6\,\micron$ data alone, the nonstellar
contribution at $5.8\,\micron$ increases modestly to $4.8\%$, while  the  
nonstellar contribution at $8\,\micron$ is not significantly changed. 
From Equation \ref{eq:sfr}, we find an upper limit to the star
formation rate in the central $1$\,kpc of NGC\,6872 of $\lesssim 0.018\sfr$.

In both the $8.0\,\micron$ total and nonstellar 
emission maps of Figure \ref{fig:n6872mosaics}, we find a 
faint stream of mid-infrared emission 
winding $\sim 4$\,kpc ($15''$) from the northwest  edge of NGC\,6872's 
ring south to the spiral galaxy's central bar and along the bar 
to NGC\,6872's nucleus (see also Fig. \ref{fig:n6872centercomp}). 
In the $8.0\,\micron$ nonstellar 
emission map, the surface brightness of the stream is a 
factor $\sim 2-3$ larger than that found elsewhere in NGC\,6872's 
central region. There is no statistically significant evidence 
for the stream feature in either the $5.8\,\micron$ total emission or 
nonstellar emission maps, where emission instead follows the stellar 
bar. The stream may be interstellar gas and dust 
that is being funneled into NGC\,6872's nuclear region by the bar. 

Numerical simulations of interacting galaxies 
often predict that gas and/or dust will be driven into the inner disk of 
the more massive galaxy, possibly inducing a circumnuclear starburst 
(see, e.g. Noguchi 1988, Hernquist \& Mihos 1995, Mihos \& Hernquist 1996).
Simulations of the NGC\,6872/IC\,4970 interaction, in isolation from its 
group environment, predict a central concentration of dense gas in the 
spiral galaxy (Mihos \etal 1993; Horellou \& Koribalski 2007) and 
$9.6 \times 10^8\Ms$ of molecular gas is observed in CO observations 
of the central $45''$ of the galaxy (Horellou \& Booth 1997).
Our {\it Spitzer} observations show that, despite the
concentration of molecular hydrogen at the center and the stream of
interstellar matter observed in $8\,\micron$ leading from the 
ring to the nucleus, there is little photometric evidence for 
PAH or warm dust emission from recent 
star formation in the nuclear region of NGC\,6872.  
This is in agreement with the  H$\alpha$ intensity maps  of the 
central region of NGC\,6872 by Mihos \etal (1993), that also indicated little 
evidence for recent nuclear or circumnuclear star formation.
However, the stream may signal the transport of interstellar matter into the 
nuclear region, as predicted by simulations, and may be a prelude to a 
future episode of nuclear star formation or AGN activity in the primary galaxy.

\begin{figure*}[t]
\begin{center}
\includegraphics[height=1.96in,width=3in]{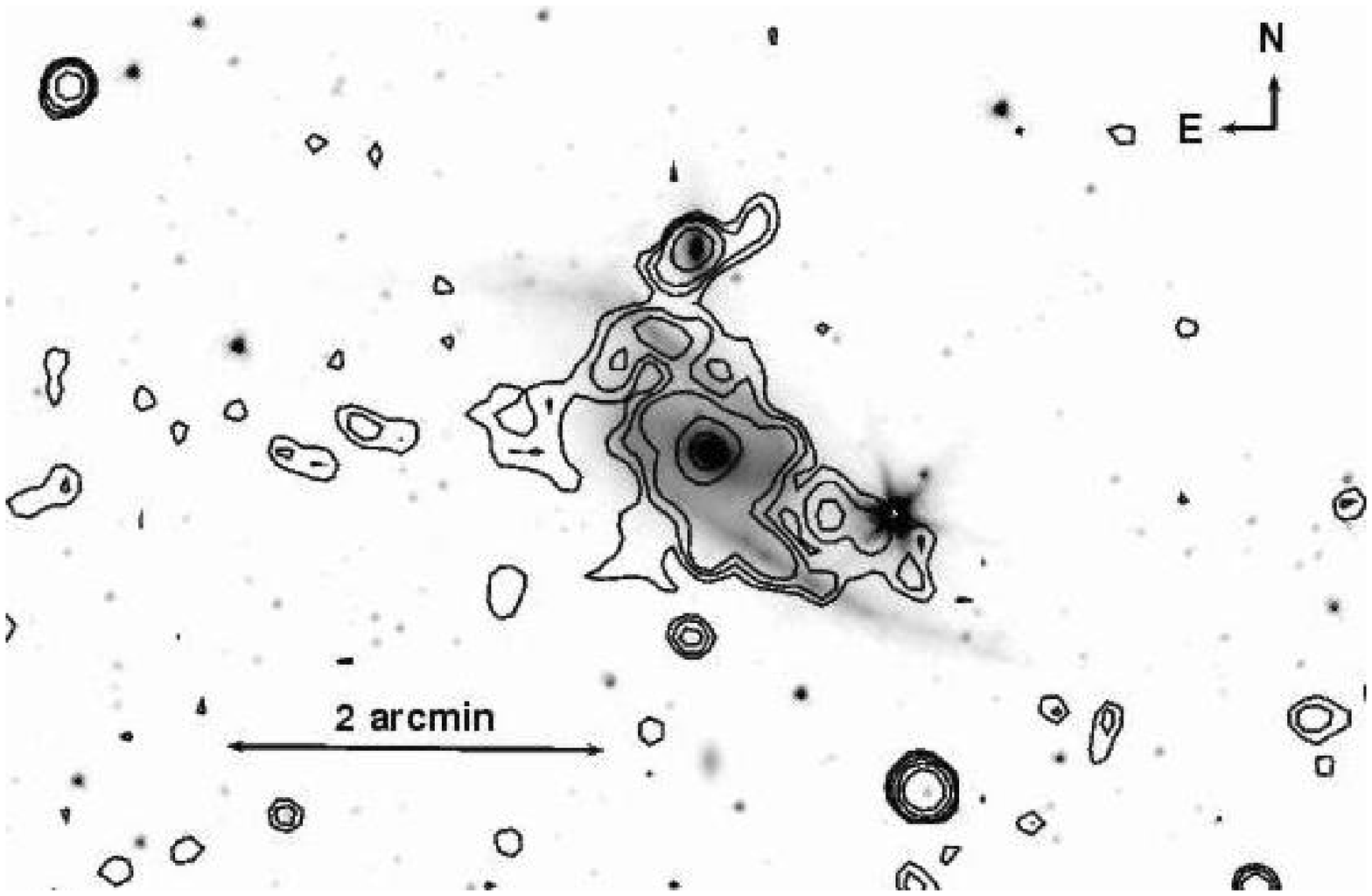}
\includegraphics[height=2.05in,width=3in]{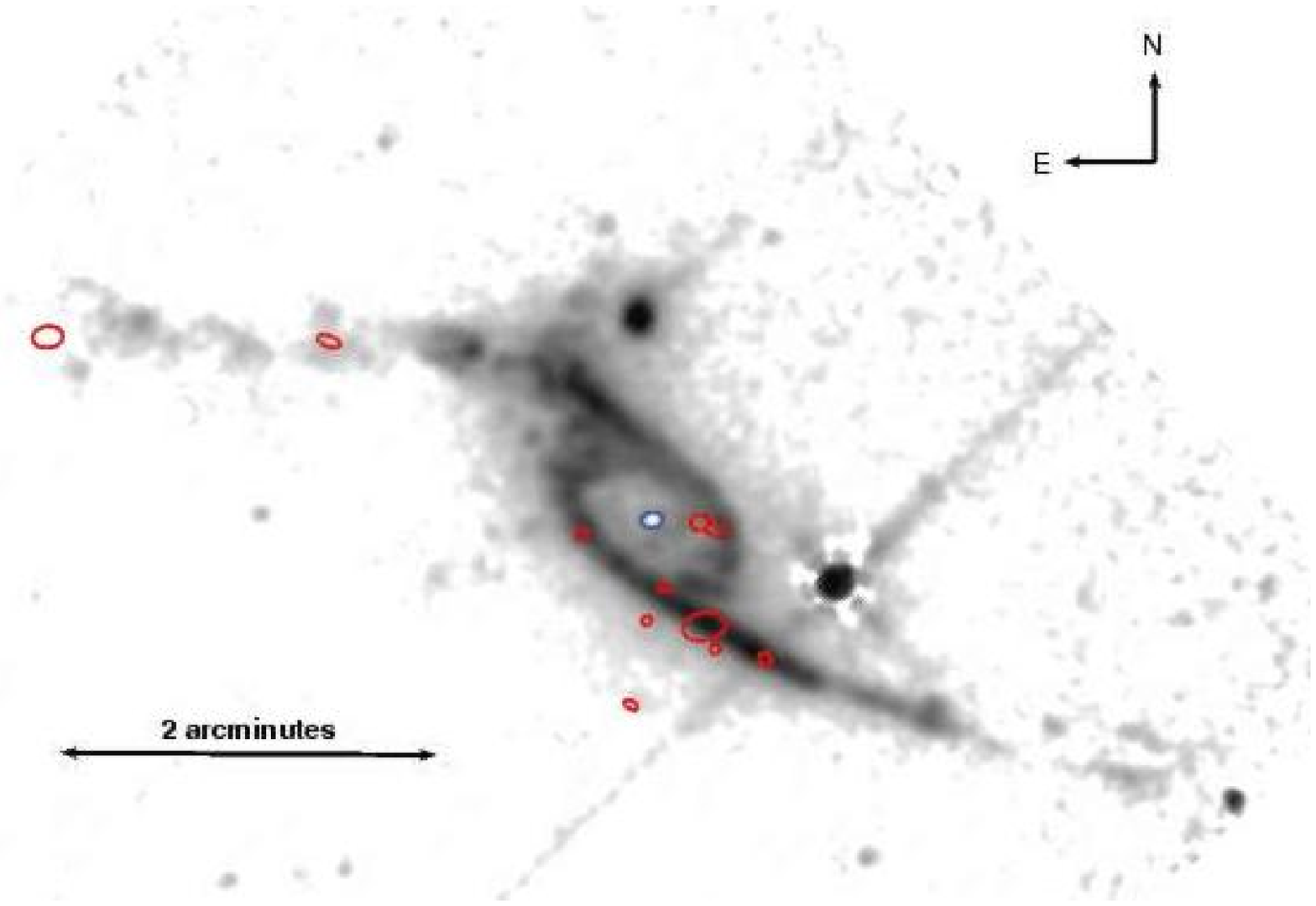}
\includegraphics[height=2.05in,width=3in]{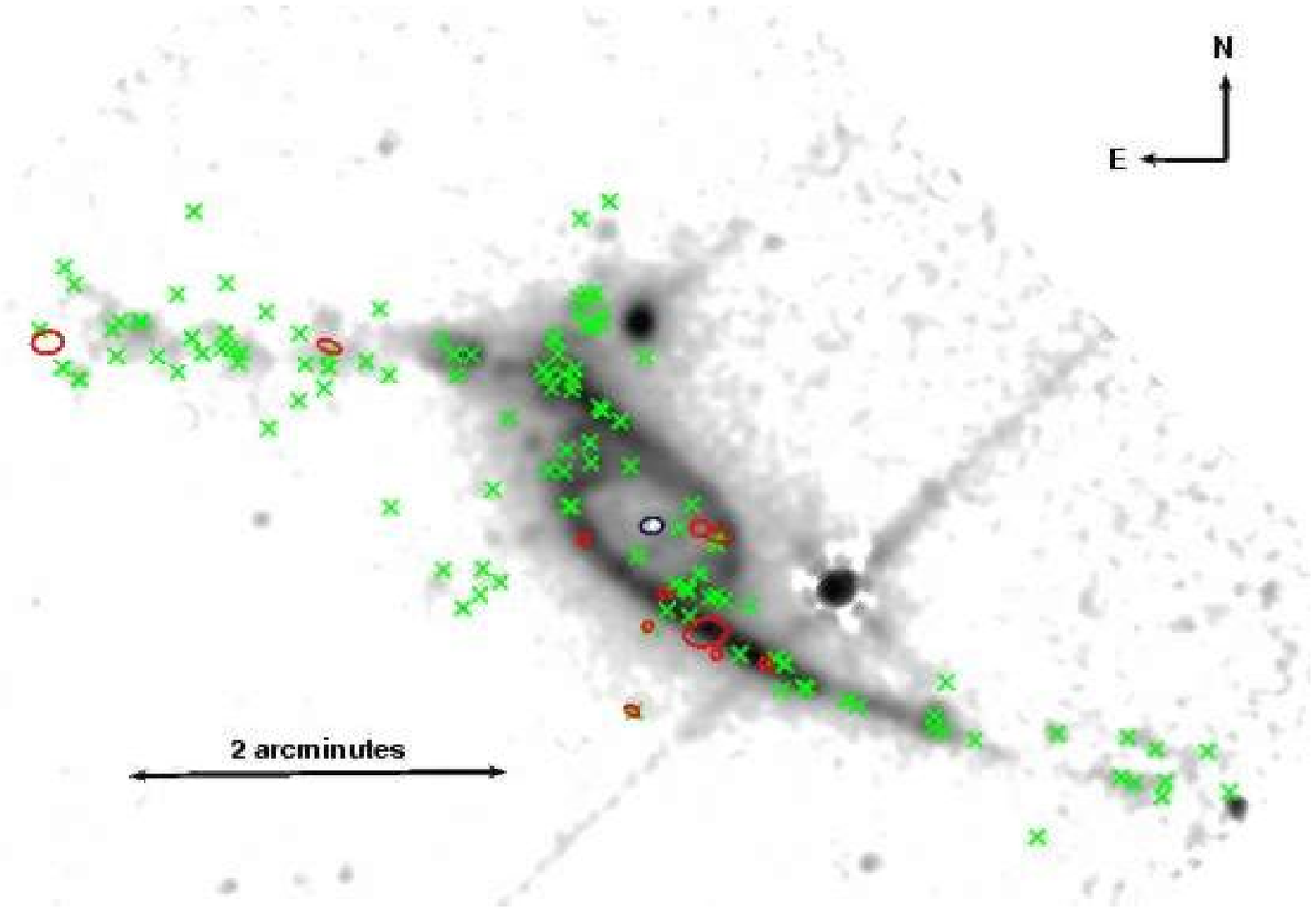}
\includegraphics[height=2.05in,width=3in]{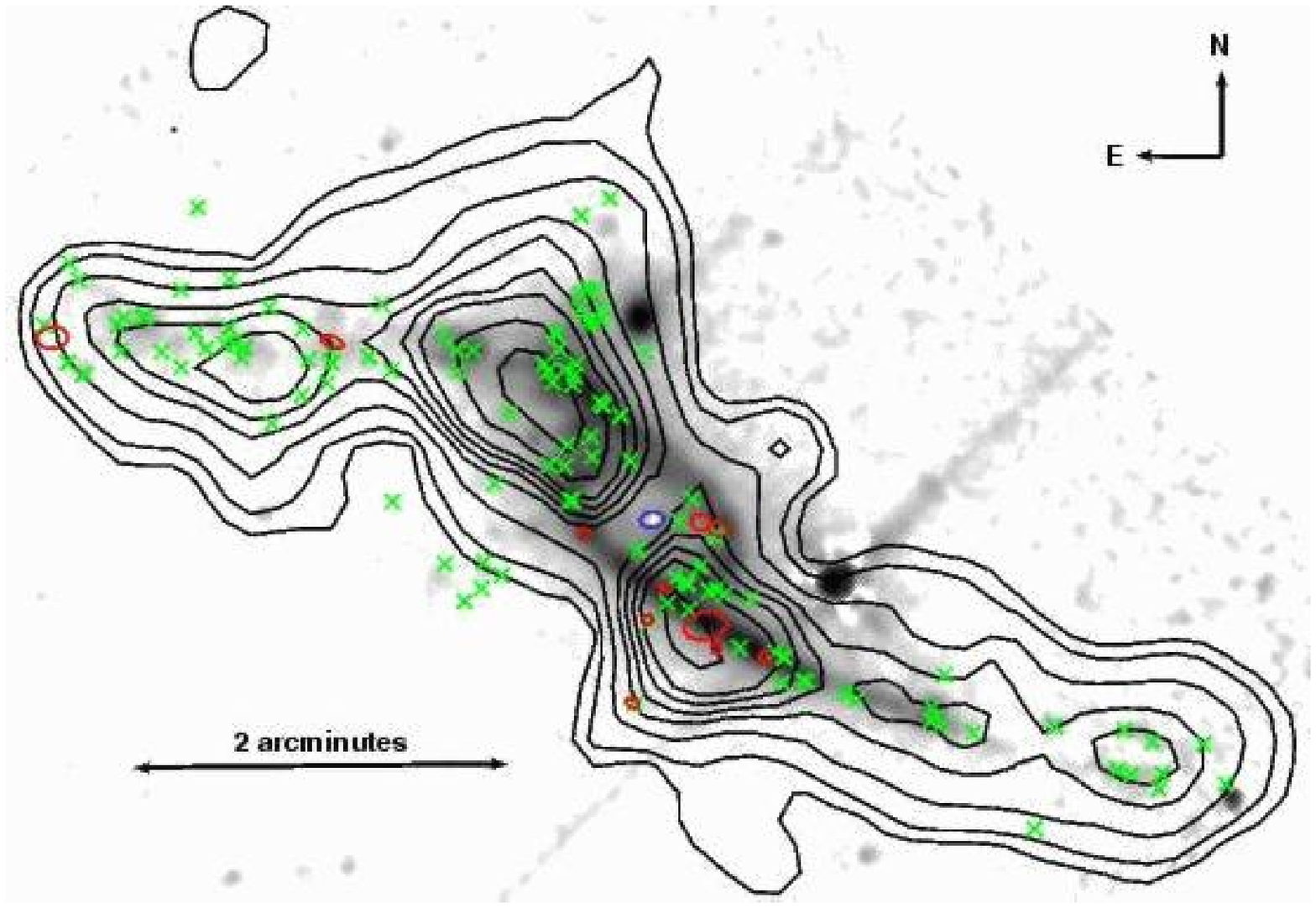}
\caption{ {\small (upper left) $3.6\,\micron$ IRAC image of the spiral 
galaxy NGC\,6872 with X-ray contours superposed. X-ray contours are
from a background subtracted, exposure corrected  $0.5-2$\,keV {\it
  Chandra} X-ray image that was smoothed with a $\sigma =4''$ Gaussian kernel.
(upper right) Nonstellar $8.0\,\micron$ IRAC image of NGC\,6872
with X-ray sources from Table \protect\ref{tab:xraysrclist}
superposed. The IRAC image has been smoothed with a 
$1\farcs5$ Gaussian kernel to highlight the faint nonstellar emission
along the full length of the tidal arms and trailing the galaxy. Red 
ellipses denote X-ray point sources, while the blue ellipse is
NGC\,6872's nucleus. The size of the ellipse for each X-ray source 
represents the $3\sigma$ wavdetect source region, set such that the 
percentage of source counts in the elliptical region corresponds to
the 3-sigma point of a corresponding two-dimensional Gaussian 
distribution, and thus depends on both the number of counts in the 
source and the detector point spread function. 
(lower left) The same $8.0\,\micron$ nonstellar image 
as the upper right panel with young star clusters (green X's) from 
Bastian \etal (2005) superposed.
(lower right) The same $8.0\,\micron$ nonstellar image as in the
lower left panel with HI contours from Horellou \& Koribalski (2007) 
superposed.
 }}
\label{fig:n6872xray}
\end{center}
\end{figure*}

\section{X-ray Comparisons to Mid-Infrared, Optical and HI in NGC\,6872}
\label{sec:xray}

\subsection{X-ray Point Sources}
\label{sec:ulxs}

Detailed studies of nearby normal spiral galaxies, e.g.  
the Milky Way (Sazonov \etal 2006; Revnivtsev \etal 2006) and 
M31 (Li \& Wang 2007; Bogdan \& Gilfanov 2008),  indicate that the 
X-ray emission from spiral galaxies is strongly correlated 
with their stellar populations, and is dominated by the
integrated emission from resolved and unresolved point sources.
In the upper left panel of Figure \ref{fig:n6872xray}, 
we superpose contours from a background-subtracted, 
exposure-corrected $0.5-2$\,keV 
Chandra X-ray image of the NGC\,6872/IC\,4970 interacting 
pair, after smoothing with a $4''$ Gaussian kernel, on the 
$3.6\,\micron$ IRAC mosaic tracing the population of old stars. 
In addition to emission from the nucleus, X-ray emission is found 
correlated with the $3.6\,\micron$ stellar map along the central 
stellar bar, in the southern spiral arm, 
and in the northern spiral arm before the break. 
In the central region of the galaxy, the X-ray emission is most likely 
the integrated emission from old stars, including coronally active low 
mass binaries, cataclysmic variable stars, and white dwarf accretors 
at the faint end of the X-ray luminosity function, as well as the more 
luminous neutron star and black hole binaries (LMXB's). 
X-ray emission from regions in  
the spiral arms may also have contributions from young stellar 
objects, young stars, and high mass X-ray binaries (HMXB's) associated 
with ongoing star formation.  

In Table \ref{tab:xraysrclist} we list the positions, observed 
X-ray source counts from the combined $70.2$\,ks {\it Chandra} 
observation,  and intrinsic $0.5-8$\.keV luminosities 
 for NGC\,6872's nucleus (N) and $11$ X-ray
point sources, in and near  the spiral galaxy.  We
expect only $1$ cosmic X-ray background source in a 
$2\farcm2 \times 1\farcm44$ rectangular region containing the main 
portion of NGC\,6872 (Brandt \etal 2001), confirming that the  
X-ray sources listed in Table \ref{tab:xraysrclist} 
are most likely associated with the spiral galaxy.  
X-ray colors for these sources using the $0.5-1$\,keV (S), $1-2$\,keV (M),
and $2-8$\,keV energy bands are also listed in Table
\ref{tab:xraysrclist} for completeness. However,
since the uncertainties in these colors are large due to the low count
rates in each band, we adopt a standard power law spectral model with 
photon index $1.6$, representative for luminous  X-ray point sources 
in galaxies (Colbert \etal 2004; Swartz \etal 2004), and Galactic hydrogen 
absorption ($5 \times 10^{20}$\cms) to calculate the intrinsic
luminosities for these sources. 
Guided by the X-ray evidence for NGC\,6872's recent passage through the
Pavo group core (Machacek \etal 2005), we adopt $55.5$\,Mpc, 
the luminosity distance to the central Pavo group galaxy NGC\,6876, 
as the luminosity distance to NGC\,6872. The $0.5-8$\,keV luminosities 
of these sources, other than the nucleus of NGC\,6872, 
range from $5 \times 10^{38}$\ergs to $5 \times 10^{39}$\ergs,
spanning the luminosities characteristic of the bright end of the 
HMXB luminosity function to those associated with ultra-luminous
X-ray sources (ULXs). 
These X-ray sources are typical of bright 
X-ray sources in other systems, where the number and luminosity of 
ultra-luminous X-ray sources are found to be correlated with galaxy 
interactions, mergers, and recent star formation (Swartz \etal 2004).

In the upper right panel of
Figure \ref{fig:n6872xray}, we overlay the $3\sigma$
wavdetect elliptical source regions (Detect Reference Manual)\footnote{
http://cxc.harvard.edu/ciao4.0/download/doc/detect\_manual/wav\_ref.html}
 for these X-ray sources on the 
IRAC $8\,\micron$ nonstellar emission map of NGC\,6872/IC\,4970. 
The $8.0\,\micron$ image has been
smoothed with a $1\farcs5$ kernel to highlight the faint nonstellar
emission extending to the ends of NGC\,6872's tidal tails. X-ray
sources other than the nucleus are found in or near star-forming
regions mapped by the nonstellar (PAH and dust) $8.0\,\micron$ emission. 
Since we have neglected intrinsic absorption, which may be 
significant in dusty star-forming regions, the X-ray  
luminosities, given in Table \ref{tab:xraysrclist}, should be 
considered lower limits on the true intrinsic X-ray luminosities of these 
sources. In the lower left panel of Figure \ref{fig:n6872xray} we
add the locations of young star clusters (green X's) in NGC\,6872 
(Bastian \etal 2005) to the $8.0\,\micron$ nonstellar image. 
Throughout the following discussion, 
all star cluster identification numbers, ages 
and masses are from Bastian \etal (2005). Please see Bastian \etal
(2005) for the full U, B, V, and I photometric results for these clusters
and a complete discussion of the star cluster age and mass modeling.
In the lower right panel of Figure \ref{fig:n6872xray}, we add 
$21$\,cm emission contours from Horellou \& Koribalski (2007), that
map the distribution of atomic hydrogen in the galaxy. The star
clusters follow closely the distribution of nonstellar $8.0\,\micron$
emission. This correspondence with $8.0\,\micron$ nonstellar
emission extends to the very ends of both the northern and 
southern tidal tails. For example, a very young star cluster
(no. $155$), with an age of only $2.5$\,Myr and mass of 
$1.2 \times 10^6\Ms$ coincides with the bright $8\,\micron$ emission knot 
found $\sim 21$\,kpc from the nucleus of NGC\,6872 
just east of the break in the northern tidal tail (see the lower right
panel of Fig. \ref{fig:n6872mosaics}), and three young ($3-6$\,Myr) 
clusters (no. $397$, $398$, and $402$) coincide with the $8\,\micron$ 
nonstellar emission double knot, $\sim 29$\,kpc from the center of the 
spiral galaxy in the southern tidal tail. Star clusters
are also found in faint $8\,\micron$ nonstellar emission clumps
trailing the galaxy to the southeast. The concentration of five star clusters   
(nos. $108$, $111$, $121$, $126$, and $130$)  located $\sim 16$\,kpc
southeast from NGC\,6872's center, as shown in the lower left panel of 
Figure \ref{fig:n6872xray}, have low to moderate masses 
($2 \times 10^4 - 6 \times 10^5\Ms$) and are also very
young, with ages between  $3.3$ to $4.6$\,Myr. These latter star clusters
may have formed from cooled gas, stripped from NGC\,6872
by tidal and turbulent-viscous forces caused by the collision of
IC\,4970 with the spiral galaxy coupled with the high velocity passage of the
NGC\,6872/IC\,4970 galaxy pair  through the Pavo group core, and may 
ultimately contribute to the Pavo intragroup stellar light. 
Young star clusters are also highly concentrated
along the bridge of nonstellar emission extending from the break in 
the northern arm to the companion galaxy IC\,4970. All but a handful of
young star clusters, as well as the bright X-ray sources and
$8.0\,\micron$ nonstellar emission, coincide with regions rich in cold HI
gas. This spatial distribution of star-forming regions and young
clusters along NGC\,6872's tidal tails and bridge connecting
IC\,4970 to NGC\,6872 strongly suggests that star formation has been
influenced, if not triggered, by the galaxies' interactions, and that the 
star clusters have only recently emerged from their dusty stellar nurseries. 

\begin{figure}[t]
\begin{center}
\includegraphics[height=2.247in,width=3in]{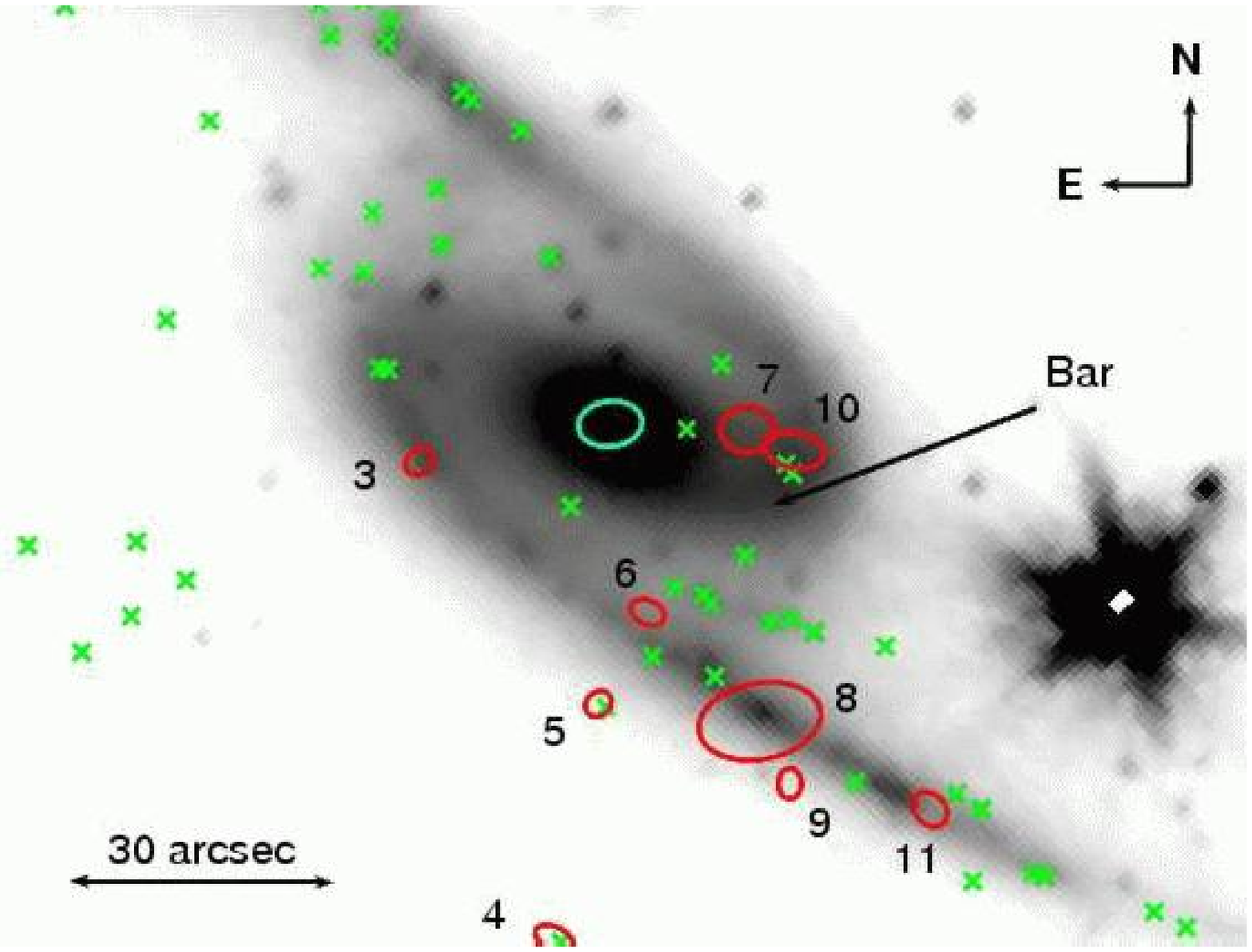}
\includegraphics[height=2.247in,width=3in]{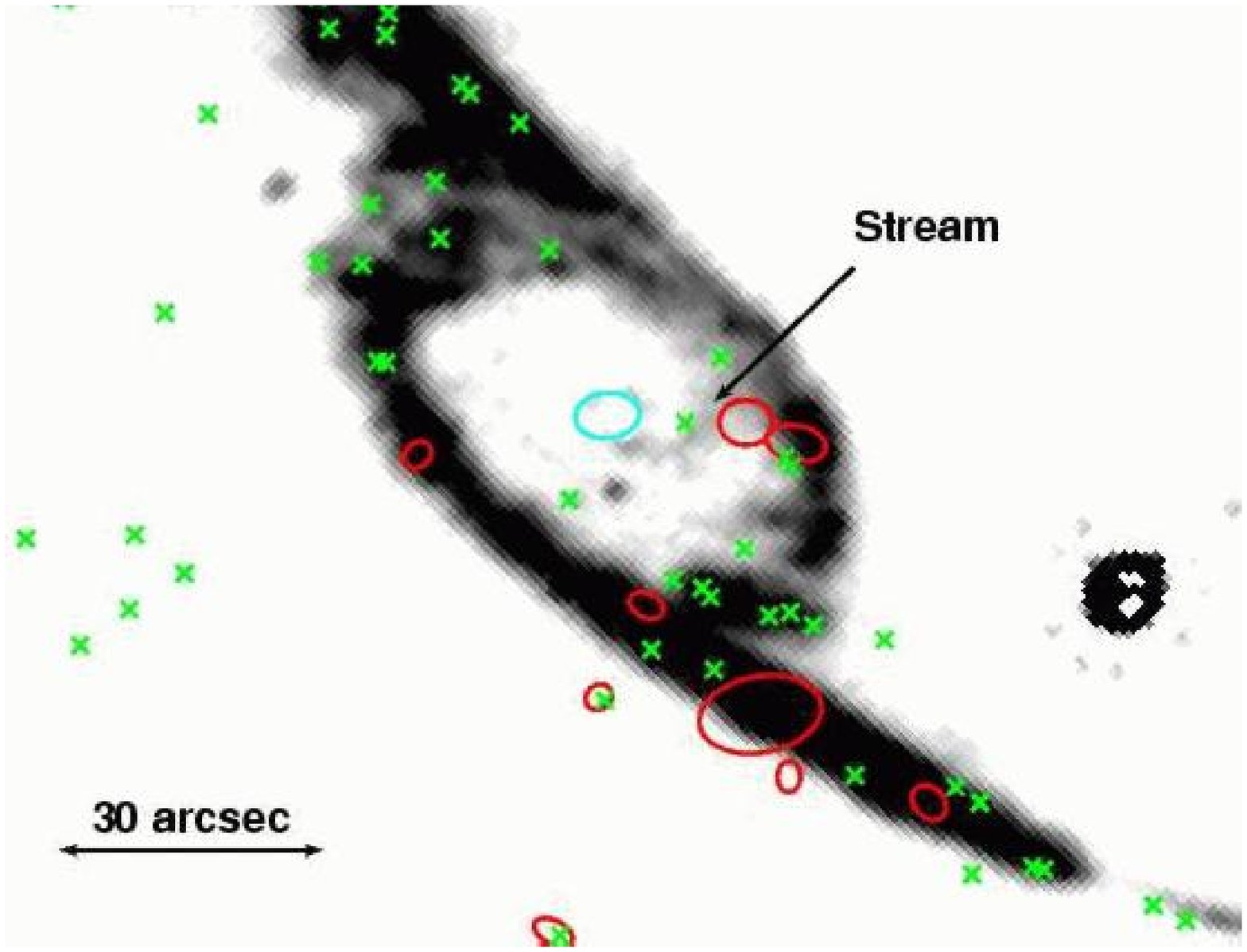}
\caption{ {\small The central region of NGC\,6872 close-up with point
 sources denoted as in Fig. \protect\ref{fig:n6872xray} 
superposed. (upper) $3.6\,\micron$ IRAC image of NGC\,6872's 
central region showing a stellar bar 
just south of X-ray sources $7$ and $10$ (see Table 
\protect\ref{tab:xraysrclist}) and strongly correlated with 
$3$ star clusters. 
(lower) The $8\,\micron$ nonstellar image of the same region
 showing the stream of emission (originating just south of a 
 star cluster in the ring and north of X-ray sources $7$ and $10$)
 winding south to the bar and then along the bar to the nucleus. 
 The image uses square root scaling. Bright regions have been 
 overexposed to highlight the faint nonstellar stream.
}}
\label{fig:n6872centercomp}
\end{center}
\end{figure}

It is interesting to note, however, that, unlike the young star
clusters,  the bright X-ray sources do not 
uniformly follow the $8.0\,\micron$ nonstellar distribution.  
While two X-ray sources are found in the northern tidal arm 
east of the break (X-ray source $1$ at the eastern tip of the arm 
and X-ray source $2$ in the middle), most of the bright X-ray sources 
are clustered to the south in NGC\,6872, and 
are seen to avoid the highly disrupted bridge region close to the 
interacting companion IC\,4970, despite the high concentration of young 
star clusters and strong nonstellar emission there. As shown in Figure
\ref{fig:n6872centercomp}, five  X-ray sources (no. $3$, $6$,
$8$, $9$, and $11$) are found in the southern spiral arm and tidal
tail. Three of these X-ray sources (no. $3$, $8$ and $11$) are
coincident or nearly coincident with the
brightest $8.0\,\micron$ nonstellar emission clumps in the southern
arm, and may be multiple 
X-ray sources associated with these compact star-forming regions. 

X-ray sources $4$  and $5$ trail the southern arm of NGC\,6872 by $\sim
 8.4$\,kpc ($32''$) and $2$\,kpc ($\sim 9''$), respectively, and  
are the only two X-ray sources coincident with optically identified 
sources (star clusters $172$ and $221$, respectively) 
from Bastian \etal (2005).  If these optical sources are star
clusters associated with the galaxy, and not foreground stars or
background AGN, the X-ray luminosities ( $3.6 \times 10^{39}$\ergs and
$5.1 \times 10^{39}$\ergs, respectively) of their associated X-ray
sources are high, characteristic of ultra-luminous X-ray sources. In 
Table \ref{tab:starclusters}, we list the  V and B band absolute
magnitudes, internal extinction factors, cluster masses and ages from 
Bastian \etal (2005) for the two star
clusters coincident with the bright X-ray sources $4$ and $5$. 
These two young, moderately massive  star 
clusters , i.e. with ages $<30$\,Myr, and masses $\sim 10^5 - 10^6\,\Ms$,
are similar in age and mass to star clusters found throughout the
tidal tails and outer parts of NGC\,6872, and in the interaction region between 
NGC\,6872 and IC\,4970 (Bastian \etal 2005). 

X-ray-to-optical flux ratios have long been used to gain insight into
the nature of X-ray emitting sources that have possible optical counterparts.
(see, e.g. Bradt \& McClintock 1983; Maccacaro \etal 1988; Stocke
\etal 1991; Hornschemeier \etal 2001 ).    
In Table \ref{tab:xor} we present three measures of the relative X-ray
to optical emission between the two ultra-luminous X-ray sources ($4$
and $5$)  and their optical hosts 
(star clusters $172$ and $221$, respectively). They are as follows: 
\begin{enumerate}
\item{the V-band X-ray-to-optical flux ratio 
(Maccacaro \etal 1982), given by 
\begin{equation}
  {\rm XOR}_V = {\rm log}(f_X) + m_V/2.5 + 5.27 
\label{eq:xorv}
\end{equation} 
where $f_X$ is the $0.3-3.5$\,keV X-ray flux (in \ergscm) and 
$m_V$ is the extinction-corrected V-band magnitude},
\item{the B-band  X-ray-to-optical flux ratio (van Paradisj \&
    McClintock 1995), defined as  
\begin{equation}
  {\rm XOR}_B = B_0 + 2.5 {\rm log}(F_{\rm X}) 
\label{eq:xorb}
\end{equation} 
where $B_0$ is the extinction corrected B magnitude and $F_{\rm X}$ is
the $2-10$\,keV X-ray flux averaged over the $2-10$\,keV
energy band (in $\mu$Jy), and}
\item{the  V-band to $1$\,keV two point spectral index, assuming a 
flux density of the form $F_\nu \propto \nu^{-\alpha ox}$}.
\end{enumerate}
We use a power law spectral model with photon index $1.6$ and Galactic
absorption to convert the observed X-ray count rates to X-ray flux in
the appropriate energy bands. From Table \ref{tab:xor} we see that
the V-band X-ray-to-optical ratio (XOR$_{\rm V} = 0.03$ ) for X-ray source $5$
is larger than expected ($> -0.5$) for any population of foreground
stars, and, although XOR$_{\rm V}=-0.9$ for X-ray 
source $4$ is compatible with the upper end of the M star range for
this ratio, the extinction corrected (B-V)$_0$ color of $-0.22$ for that
source is too blue to be an M star (Stocke \etal 1991) . Thus neither 
source is a foreground star. The V-band X-ray-to-optical ratios for
the optical counterparts of X-ray sources $4$ and $5$ lie in the range 
($-1.5$ to $0.7$) observed for luminous young star clusters hosting 
ULXs in NGC\,4038/NGC\,4039 (Zezas \etal 2002) and NGC\,7714/NGC\,7715 
(Smith \etal 2005b). Similarly, the V-band absolute magnitude 
($-14.3$) of star cluster $172$ (associated with X-ray source $4$) is 
similar  to the absolute  V magnitude ($-14.1$) of the near-nuclear 
young star cluster hosting a bright X-ray source in interacting galaxy 
NGC\,7714 (Smith \etal 2005b), and the absolute  V magnitude ($-11.7$) of star
cluster $221$ (associated with X-ray source $5$) is comparable to 
absolute V magnitudes ($-10$ to $-13.7$) of star clusters associated
with  ULXs in NGC\,4038/NGC\,4039 (Zezas \etal 2002). Also the 
V-I colors for star clusters $172$ and $221$ 
are blue, $0.74$ and $0.86$, respectively (Bastian \etal 2005),  
similar to colors of star-forming dwarf galaxies in other nearby loose groups 
(Carrasco, de Oliveira \& Infante 2006). The B-band X-ray-to-optical ratios
($11.2$ and $13.9$ for sources $4$ and $5$, respectively) are similar 
to those observed for  HMXBs (van Paradijs \& McClintock 1995). However,  
since stars or accreting binaries are optically too faint
to be individually detected  at the luminosity distance of NGC\,6872, 
these B-band X-ray to optical ratios likely reflect the presence of 
HMXBs in and the young age of the X-ray sources'  star cluster hosts. 
Taken together, these properties suggest that X-ray sources 
$4$ and $5$ reside in young star-forming clusters associated with the 
interacting galaxy NGC\,6872 in the Pavo group (Bastian \etal 2005).  
However, since the V-band 
X-ray-to-optical ratios and the two-point V-band to $1$ keV spectral
indices are also consistent with those observed from AGN (Stocke \etal
1991, Landt \etal 2001), we can not rule out the possibility that one
or both of these sources may be background AGN. Optical 
spectra and redshifts are needed to determine whether these X-ray sources
are, indeed, ULXs associated with the NGC\,6872/IC\,4970 interacting
galaxies in the Pavo group. 

Two X-ray sources (sources $7$ and $10$) are found west of NGC\,6872's
nucleus on the western edge of the ring just north of the stellar bar (see upper
panel of Fig. \ref{fig:n6872centercomp}). In the lower panel of Figure 
\ref{fig:n6872centercomp}, we see that the two western X-ray sources 
lie at the base of the stream of interstellar matter, seen in 
$8\,\micron$ nonstellar emission, that may be feeding the nucleus. 
Four star clusters (green X's ) also are observed in this 
region: star cluster $320$ to the northwest at the inner edge of the ring,
star clusters $331$ and $338$ displaced $\lesssim 0.7$\,kpc from  
 X-ray source $10$, and star cluster $301$ in the bulge. The northwest 
and bulge star clusters trace the northern edge of the stream, while X-ray 
sources $7$ and $10$ and nearby star clusters lie at the stream's
southern base. The two star clusters close to X-ray source $10$, 
as well as the star cluster located  to the northwest just inside the 
$8\,\micron$ emission ring, are more massive ($\sim 10^7\Ms$) and older 
($\sim 1$\,Gyr) than most of the star clusters in the tidal
features. However, the inner star cluster near the bulge on the
northern edge of the stream is young ($\sim 3$\,Myr).  Stellar winds
and supernovae from these clusters, along with existing interstellar material
and dust funneled to the center by the bar, may
contribute to the warm dust and gas found in the stream.

\subsection{Relationship of Diffuse X-ray Emission to Star Formation}  
\label{sec:diffuse}

\begin{figure}[t]
\begin{center}
\includegraphics[height=1.96in,width=3in]{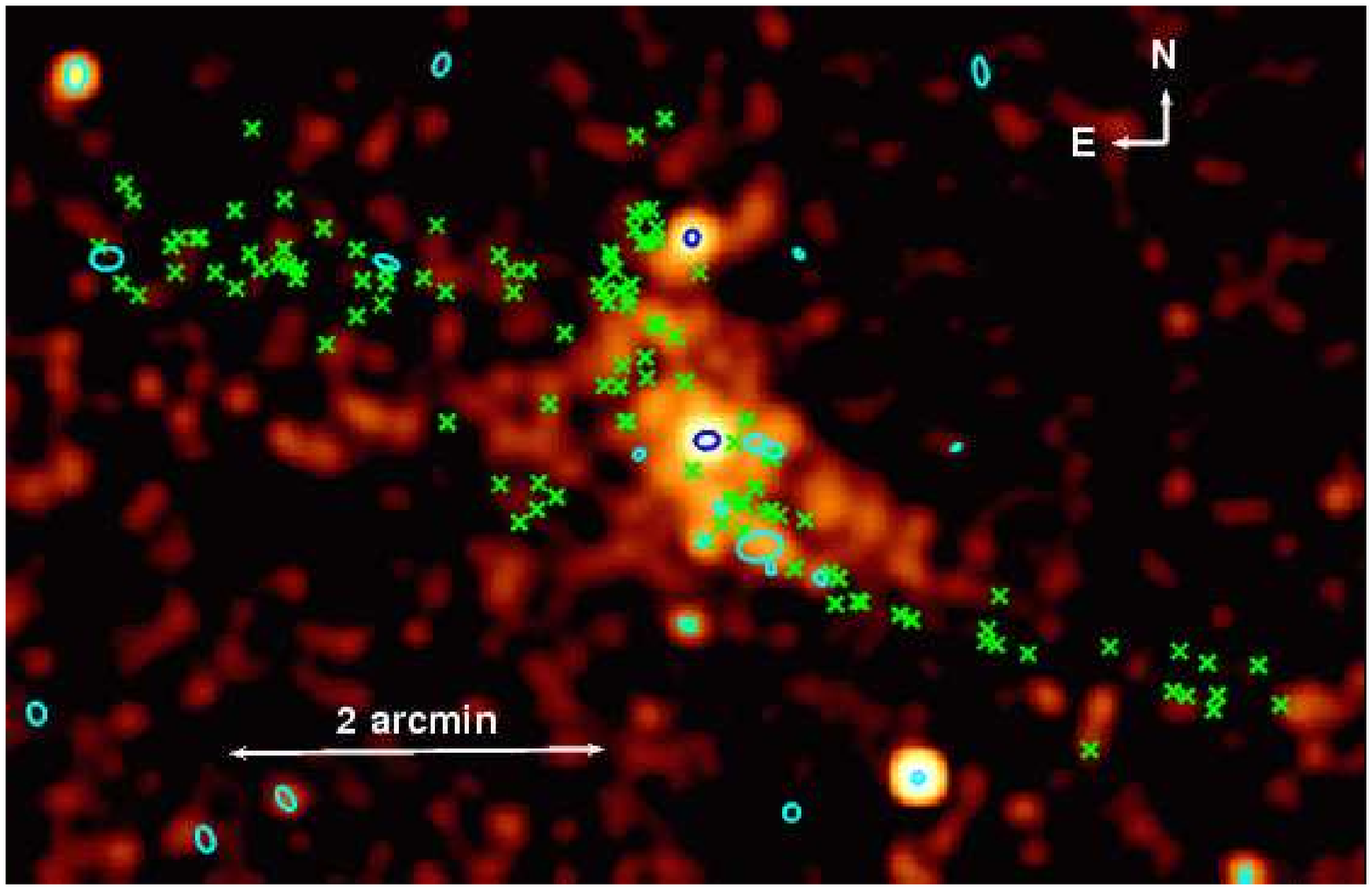}
\includegraphics[height=2.46in,width=3in]{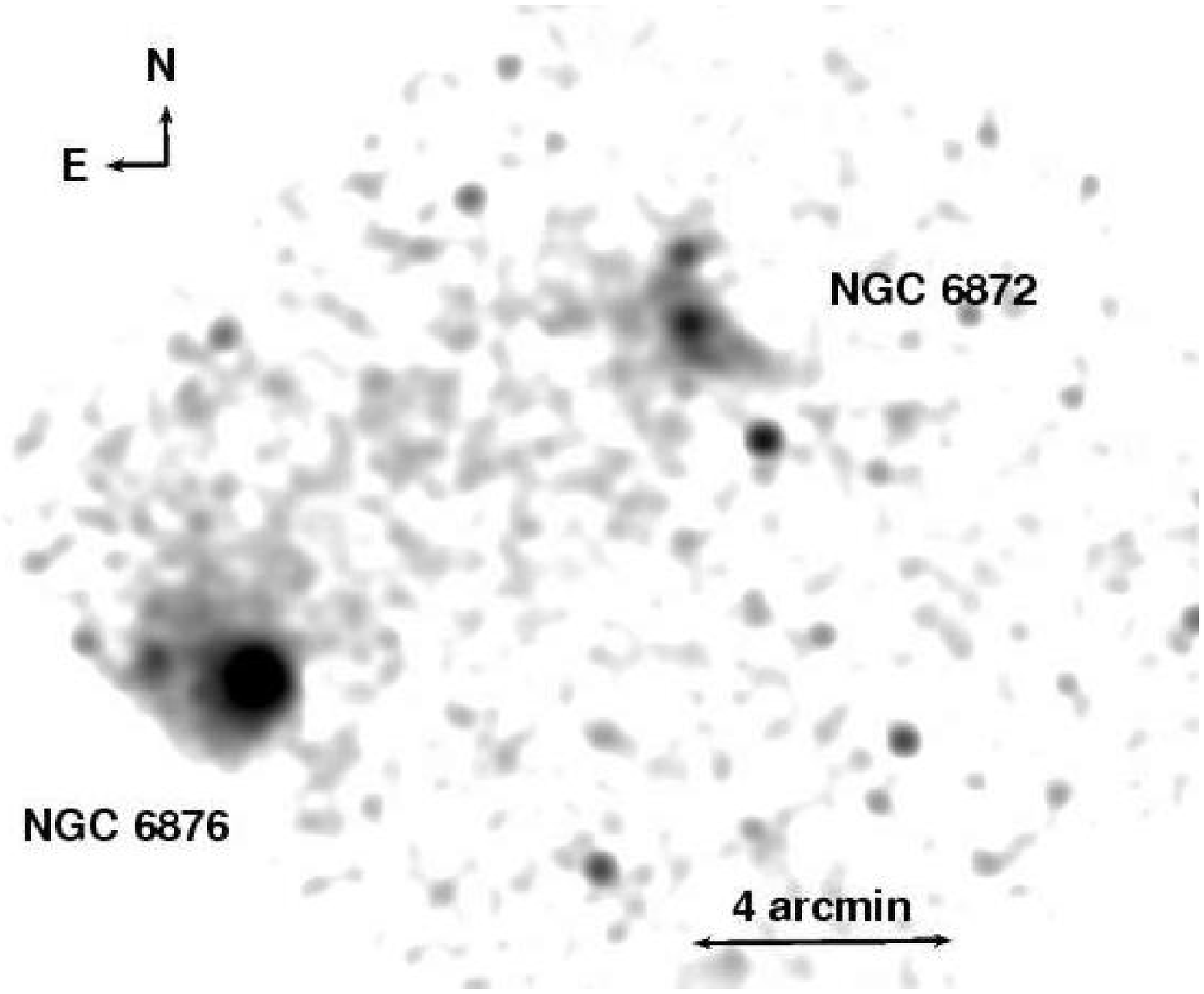}
\caption{{\small (upper) $0.5-2$\,keV X-ray image of NGC\,6872
  (see upper left  panel of Fig. \protect\ref{fig:n6872xray} 
  with X-ray point sources
  (ellipses as in Fig. \protect\ref{fig:n6872xray}) and 
  young star clusters (X's) superposed. The X-ray image
  is background-subtracted, exposure corrected and smoothed with a 
  $\sigma=4''$ Gaussian. 
(lower)  $0.5-2$\,keV {\it Chandra} X-ray image of the
  core of the Pavo group, confirming the excess X-ray emission 
 (trail), found in 
{\it XMM-Newton} images (Machacek \etal 2005b), between the 
 dominant elliptical galaxy NGC\,6876 and the spiral galaxy NGC\,6872.
The {\it Chandra} image has been background subtracted, exposure
corrected and smoothed with an $8''$ Gaussian kernel.
}}
\label{fig:nonstellarcompare}
\end{center}
\end{figure}

The upper left panel of Figure \ref{fig:n6872xray} shows X-ray emission that 
deviates strongly from the underlying distribution of stars in
NGC\,6872. Diffuse X-rays in normal, non-interacting 
spiral galaxies are expected to arise from two principal sources:  
a soft ($\sim 0.3-0.4$\,keV) thermal component from winds of 
ionized gas ejected from evolved stars and Type Ia supernovae and 
a hard component from core collapse supernovae, associated with 
star-forming regions (Bogdan \& Gilfanov 2008). Since 
   these components are associated with
outflows, they need not be correlated with the stellar distribution. 
For interacting spiral galaxies, gas may be stripped and/or heated by 
the interaction itself, and, especially when the interacting partner
is an early type galaxy such as IC\,4970, hot gas may be stripped from the 
hot halo of the interacting companion into the space 
between the galaxies and/or transferred onto the massive spiral galaxy.    
Thus the presence of diffuse hot gas with disturbed
morphologies in the NGC\,6872/IC\,4970 interacting pair is not a surprise.

In the lower panel of Figure \ref{fig:nonstellarcompare} we show the 
background subtracted, exposure corrected $0.5-2$\,keV {\it Chandra} 
image of X-ray emission in the NGC\,6872/IC\,4970 interacting galaxy
pair, after smoothing with a $4''$ Gaussian kernel. As in
Fig. \ref{fig:n6872xray} we overlay the positions of young star clusters 
as green X's and X-ray point sources as ellipses. 
We see bright X-ray emission
and a sharp surface brightness edge leading the stellar distribution
and young star clusters in NGC\,6872's southern spiral arm. X-ray
emission is truncated just west of the stellar tidal bridge leading
from NGC\,6872's northern spiral arm to IC\,4970, and then turns 
sharply in a bright $\sim 20$\,kpc long filament to the southeast 
in the direction of NGC\,6876. Faint diffuse emission is also seen 
extending $\sim 16$\,kpc behind the galaxy to the southeast of 
NGC\,6872's nucleus. While young star clusters are strongly correlated
with the nonstellar $8.0\,\micron$ emission in NGC\,6872 and trace the
tidally distorted spiral arms and bridge to IC\,4970 
(see Fig. \ref{fig:n6872xray}), they avoid the bright regions of hot
diffuse X-ray gas, and are found instead bordering the diffuse 
X-ray emission in the X-ray filaments and the fainter patches of X-ray
emission in the tidal arms and trailing NGC\,6872 to the
southeast. This suggests that star clusters formed in the tidal debris 
from gas, previously stripped by the interactions
of the spiral galaxy with IC\,4970 and the Pavo group, that has since
cooled.  These clusters  may ultimately contribute to the Pavo group's
intergalactic stellar light.  

In this paper we have limited our discussion of the diffuse X-ray gas
to comparisons with star formation tracers in NGC\,6872. However,  the
hot gas to the east in NGC\,6872, is highly disturbed, bending  
sharply to the south, away from the eastern tidal tail, and extending 
to the southeast of NGC\,6872's stellar distribution until it merges 
into the X-ray trail (see Fig. \ref{fig:nonstellarcompare} and Machacek \etal
2005). This disturbed morphology signals the importance of 
hydrodynamical forces, such as ram 
pressure or turbulent viscosity, that are  caused by high velocity 
interactions between the galaxies and the ambient group IGM. These
interactions act strongly on the diffuse hot galaxy gas, but are 
ineffective at disturbing the stars. We discuss the detailed
thermodynamic properties of the hot gas 
in and around the dominant Pavo galaxies (NGC 6876 and NGC 6872), 
the signatures of the supersonic passage of NGC 6872 through the core 
of the Pavo group that are imprinted on that gas, and the impact of 
these galaxy-group interactions on the hydrodynamical state of the Pavo IGM 
in a forthcoming paper (Machacek \etal 2008b, in preparation).

\section{Conclusions}
\label{sec:conclude}

The dominant galaxies in the Pavo galaxy group core, i.e. the large spiral 
galaxy NGC\,6872, central group elliptical NGC\,6876, and their
companion galaxies IC\,4970 and NGC\,6877,  provide an ideal nearby 
laboratory to study the combined effects of galaxy-galaxy tidal 
interactions and high velocity galaxy-IGM interactions on star
formation and galaxy evolution. In this paper we present results from  
{\it Spitzer} mid-infrared observations of these galaxies in the 
$3.6$, $4.5$, $5.8$, and $8.0\,\micron$ IRAC bands, and {\it Chandra} X-ray 
observations of the X-ray point source population and hot gas in 
the spiral galaxy NGC\,6872. We combined these results with archival 
optical and HI data to study the properties of interaction-induced 
star formation in and near the large spiral galaxy. 
We find the following:
\begin{itemize}

\item{Excess $8\,\micron$ emission in NGC\,6876, the central Pavo 
 group elliptical galaxy, 
 is weak and follows the distribution of starlight. 
In the central $1.6$\,kpc ($6\farcs1$) of NGC\,6876, $\sim 16\%$ of 
the $8.0\,\micron$ emission is `nonstellar' 
(as defined in \S\ref{sec:irac} and Pahre \etal 2004), 
while we find no statistically significant nonstellar 
emission at $5.8\,\micron$. This suggests that the nonstellar 
$8.0\,\micron$ flux may reflect  leakage of the broad $9.7\,\micron$ 
emission feature, produced by warm silicate dust grains ejected  from 
the atmospheres of evolved AGB stars,  into the $8.0\,\micron$
bandpass.}

\item{Nonstellar emission, with mid-infrared colors consistent with 
emission from PAH molecules and a warm dust continuum, 
contributes $48\%$ ($14\%$) of the 
total  $8.0\,\micron$ ($5.8\,\micron$) emission in the central $1$\,kpc 
of NGC\,6877, the elliptical companion galaxy to NGC\,6876, and suggests 
the presence of warm dust irradiated by recent star formation in
NGC\,6877's nuclear region.
}

\item{In the spiral galaxy NGC\,6872, $8.0\,\micron$ and $5.8\,\micron$ 
nonstellar emission, with mid-infrared colors consistent with emission
from PAH molecules and warm dust, is found concentrated in 
clumps in a $5$\,kpc radius outer ring about the center of the spiral galaxy, 
in a bridge of emission connecting NGC\,6872's northern spiral arm 
to IC\,4970, and $\sim 50$\,kpc  northeast and 
southwest of the nucleus along the full extent of NGC\,6872's tidal
arms. 
}

\item{The mid-infrared colors in NGC\,6872 are consistent with those
  found in star-forming regions in other interacting galaxies. The 
$S_{3.6}/S_{4.5}$ flux ratio ($ \sim 1.4$) in the bright nonstellar 
 emission regions is depressed relative to elsewhere in the galaxy. 
The $[4.5]-[5.8]$ color reddens from $\sim 0.1-0.2$ in the inner
  $5$\,kpc of the galaxy, where there is little nonstellar emission, 
to $0.84$ in the ring  and   $\gtrsim 1.7$ in the bright nonstellar 
emission clumps in
  the bridge and the tidal arms, indicating a younger 
stellar population and/or more interstellar dust in these features. 
The $S_{8.0}/S_{3.6}$ flux ratio, a measure of the star formation rate
  per unit mass, is as much as a factor $\sim 20$ higher in the
  bright nonstellar emission clumps in the tidal features in NGC\,6872 
  than in the  spiral galaxy's center. The $[5.8]-[8.0]$ color 
  of the nonstellar emission in the bright clumps is $2.1$, in 
  excellent agreement with the theoretical prediction for 
  emission from PAH emission and warm dust. } 

\item{We find no strong differences in the mid-infrared colors of the 
star-forming regions in the spiral galaxy NGC\,6872 as a function of
their position relative to the tidally interacting companion galaxy IC\,4970.}

\item{Young star clusters are strongly correlated with the regions of 
intense nonstellar emission in the bridge and tidal arms. 
The brightest nonstellar emission clumps and most of the star clusters 
also correspond to regions rich in HI gas. }

\item{Eleven X-ray sources with luminosities  
 $\gtrsim (0.5 - 5)\times 10^{39}$\ergs are found 
strongly correlated with  non-stellar $8\,\micron$ emission and young 
star clusters in NGC\,6872, suggesting that these 
sources, while near ULX luminosities, may be bright HMXB's.
 Most of these bright sources are clustered to the south in NGC\,6872, 
 and avoid the highly disrupted bridge region close to the tidally 
interacting companion IC\,4970, despite a high concentration of young
 star clusters and strong nonstellar $8.0\,\micron$ emission in that region. 
}

{\item The nucleus of NGC\,6872 is a weak X-ray point source with 
$0.5-8$\,keV luminosity of $8.5 \times 10^{39}$\ergs.
Mid-infrared colors of the nuclear region of NGC\,6872 
are consistent with normal (nonactive) galaxies dominated by
starlight. Statistically significant ($23\%$ or $0.84$\,mJy)
nonstellar emission is observed at $8.0\,\micron$, but not in the 
$5.8\,\micron$ ($\lesssim 4.8\%$) bandpass. Thus, while dust is present,
there is little evidence in the inner $1$\,kpc ($3\farcs66$) of
NGC\,6872  for PAH emission from recent star formation or nuclear activity.
}  

\item{We observe a $4$\,kpc ($15''$) stream of interstellar matter at
$8$\,micron, that extends from NGC\,6872's ring to the galaxy's
nuclear region and is bordered by three young star clusters 
and two X-ray sources.
 The $8.0\,\micron$ 
 nonstellar surface brightness in the stream is a factor $2-3$
  greater than in the surrounding central region of NGC\,6872. 
 This stream may signal the transport of matter along NGC\,6872's
  central bar into the nucleus, as predicted
  by simulations, and may be a prelude to a future nuclear starburst
  or episode of strong AGN activity in the primary galaxy.
}

\item{  Mid-infrared nonstellar (PAH and dust) emission clumps and 
  young star clusters are found in NGC\,6872's bridge, tidal arms and   
  trailing the spiral galaxy to the southeast, 
  bordering regions of diffuse X-ray gas. This may suggest that stars 
  form as gas stripped by the interaction cools. } 

\end{itemize}


\acknowledgements
Support for this work was provided, in part, by the National Aeronautics
and Space Administration (NASA) through {\it Chandra} Award Number
GO6-7068X issued by the {\it Chandra} X-ray Observatory Center, which is
operated by the Smithsonian Astrophysical Observatory for and on
behalf of NASA under contract NAS8-03060, by NASA through an award
issued by JPL/Caltech, and by the Smithsonian Institution. This work
is based in part on observations made with 
the Spitzer Space Telescope, which is operated by the Jet Propulsion 
Laboratory, California Institute of Technology under a contract with NASA. 
This work made use of  the IRAC post-BCD processing software 
"IRACProc" developed by Mike Schuster, Massimo Marengo and Brian Patten at 
the Smithsonian Astrophysical Observatory, and also 
data products from the Two Micron All Sky Survey, which is a joint 
project of the University of 
Massachusetts and the Infrared Data Analysis and Processing 
Center/California Institute of Technology, funded by NASA
and the National Science Foundation, and has used the NASA/IPAC 
Extragalactic Database (NED), which is operated by JPL/Caltech, 
under contract with NASA has also been used.  
 We especially thank Massimo Marengo for help with IRACProc, 
 and Cathy Horellou and B\"arbel Koribalski for use of their HI maps of 
 NGC\,6872. 

\noindent{\it Facilities:} CXO (ACIS-I), Spitzer (IRAC)


\begin{deluxetable}{lccc}
\tablewidth{0pc}
\tablecaption{NGC\,6876 Mid-Infrared Photometry\label{tab:n6876phot}}
\tablehead{\colhead{Region} &\colhead{Waveband} & \colhead{Flux density} &
   \colhead{Magnitude} \\
 & ($\,\micron$) & (mJy) &  }
\startdata
Full &        &  &     \\
     & $3.6$  &$136.3$  & $8.28$  \\
     & $4.5$  &$81.5$  & $8.36$    \\
     & $5.8$  &$61.1$  & $8.19$    \\
     & $8.0$  &$33.2$  & $8.21$   \\
Full NS &   &   &  \\
     & $5.8$ & $5.38$  & \ldots  \\  
     & $8.0$ & $0.99$  & \ldots \\
Center &     &   &      \\
     & $3.6$  &$17.34$  &$10.52$      \\
     & $4.5$  &$9.96$  & $10.64$    \\
     & $5.8$  &$7.52$  & $10.46$   \\
     & $8.0$  &$4.78$  & $10.32$    \\
Center NS &  &  &  \\
     & $5.8$  &$0.57$  & \ldots    \\
     & $8.0$  &$0.77$  & $12.31$    \\
\enddata
\tablecomments{\small{Apertures used are (1) Full: an ellipse with 
  semi-major(minor) axes of $40\farcs7$($36\farcs0$) and position 
  angle $177^\circ$ with a concentric elliptical background annulus
  with outer and inner semi-major (minor) axes of $60\farcs0$($53\farcs1$)
  and $40\farcs7$($36\farcs0$), respectively; (2) Center: a $6\farcs1$ 
  circular aperture with a concentric circular
  background annulus with  (inner, outer) radii of 
  ($6\farcs1$,$12\farcs2$); (3)  
  Full NS and Center NS: the same aperture as (1) and (2),
  respectively, with stellar model subtracted. 
  All apertures are centered on the 
  nucleus of NGC\,6876 ($20^h18^m19^s.2$, $-70^\circ51'\,31\farcs7$) 
  determined from the peak in the $3.6\,\micron$
  emission. Uncertainties in the flux density measurement are $\sim 10\%$}
  }
\end{deluxetable}

\begin{deluxetable}{lccc}
\tablewidth{0pc}
\tablecaption{NGC\,6877 Mid-Infrared Photometry\label{tab:n6877phot}}
\tablehead{\colhead{Region} &\colhead{Waveband} & \colhead{Flux density} &
   \colhead{Magnitude} \\
 & ($\,\micron$) & (mJy) &  }
\startdata
Full &          &   &  \\
     & $3.6$ &$32.0$ &$9.86$    \\
     & $4.5$ &$19.6$ &$9.91$   \\
     & $5.8$ &$14.7$ &$9.73$   \\
     & $8.0$ &$9.32$  &$9.59$   \\
Full NS &  &  &  \\
     & $5.8$  &$1.47$ & $12.23$  \\
     & $8.0$  &$1.66$  & $11.46$  \\
Center &     &   &      \\
     & $3.6$  &$8.62$  & $11.28$      \\
     & $4.5$  &$5.18$  & $11.34$    \\
     & $5.8$  &$4.12$  & $11.11$    \\
     & $8.0$  &$3.86$  & $10.55$    \\
Center NS & & & \\
     & $5.8$  &$0.59$  &$13.23$     \\
     & $8.0$  &$1.81$  & $11.37$     \\
\enddata
\tablecomments{\small{Apertures used are (1) Full: an ellipse with 
  semi-major(minor) axes of $21\farcs4$($12\farcs7$) and position 
  angle $256^\circ$ with a concentric elliptical background annulus
  with outer and inner semi-major (minor) axes of $40\farcs$($23\farcs74$)
  and $21\farcs7$($12\farcs7$), respectively; (2) Center: a $3\farcs66$ 
  circular aperture with a concentric circular
  background annulus with  (inner, outer) radii of 
  ($3\farcs66$,$8\farcs54$); (3) Full NS and Center NS:  the same
     aperture as (1) and (2), respectively, 
  with stellar model subtracted. All apertures are centered on the 
  nucleus of NGC\,6877 ($20^h18^m36^s.1$, $-70^\circ51'\,12\farcs0$)
  determined from the peak in the $3.6\,\micron$ emission. 
  Uncertainties in the flux density measurements are $\sim 10\%$}
  }
\end{deluxetable}

\begin{deluxetable}{lccc}
\tablewidth{0pc}
\tablecaption{Mid-infrared Colors of NGC\,6876 and NGC\,6877\label{tab:n6876colors}}
\tablehead{\colhead{Region} &\colhead{$[3.6]-[4.5]$} &
  \colhead{$[4.5]-[5.8]$} & \colhead{$[5.8]-[8.0]$} }
\startdata
NGC\,6876 &  & & \\
\,\,\,Full & $-0.07$ & $0.17$  & $-0.03$  \\
\,\,\,Center & $-0.12$ & $0.18$ & $0.20$ \\
NGC\,6877 & & & \\
\,\,\,Full & $-0.05$ & $0.18$ & $0.14$ \\
\,\,\,Center & $-0.06$ & $0.56$ & $0.73$ \\
\enddata
\tablecomments{\small{ Regions for NGC\,6876 and NGC\,6877 are defined
in Tables \protect\ref{tab:n6876phot}  and \protect\ref{tab:n6877phot}, 
respectively. Colors are given in magnitudes.}
}
\end{deluxetable}

\begin{deluxetable}{lccc}
\tablewidth{0pc}
\tablecaption{Mid-Infrared Photometry of NGC\,6872's Nucleus\label{tab:n6872phot}}
\tablehead{\colhead{Component} &\colhead{Waveband} & \colhead{Flux density} &
   \colhead{Magnitude} \\
 & ($\,\micron$) & (mJy) &  }
\startdata
All  &          &   &  \\
     & $3.6$ &$12.3$ &$10.90$    \\
     & $4.5$ &$7.11$  &$11.01$   \\
     & $5.8$ &$5.10$  &$10.88$   \\
     & $8.0$ &$3.65$  &$10.61$   \\
Nonstellar & & & \\
     & $5.8$ &$0.15$ & \ldots\\
     & $8.0$ &$0.79$ & $12.27$ \\
\enddata
\tablecomments{\small{Mid-infrared fluxes for the nucleus of NGC\,6872 
   measured in a $3\farcs66$ circular aperture with a concentric circular
  background annulus whose (inner, outer) radii are 
  ($3\farcs66$,$8\farcs54$). `All' denotes the total flux (stellar +
  nonstellar), while nonstellar denotes the flux with the stellar
  model subtracted.  The apertures were centered on the 
  nucleus of NGC\,6872 ($20^h16^m56.46s$, $-70^\circ46'4\farcs80$, J2000)
  determined from the peak in the $3.6\,\micron$
  emission. Uncertainties in the flux density measurements are $\sim 10\%$}
  }
\end{deluxetable}

\begin{deluxetable}{ccc}
\tablewidth{0pc}
\tablecaption{Mid-Infrared Colors for NGC\,6872's Nucleus\label{tab:n6872col}}
\tablehead{\colhead{Flux Ratio} & \colhead{Photometry} & \colhead{Map} \\
    Color & & } 
\startdata
$S_{3.6}/S_{4.5}$ &  $1.73$  & $1.75$   \\
$[3.6]-[4.5]$ & $-0.11$& $-0.12$\\
 & & \\
$S_{5.8}/S_{4.5}$ & $0.72$ & $0.70$\\
$[4.5]-[5.8]$  &$0.12$  & $0.10$  \\
 & & \\
$S_{8.0}/S_{3.6}$ & $0.30$ & $0.28$ \\
$[3.6]-[8.0]$  &$0.29$  &$0.23$   \\
 & & \\
$S_{8.0}/S_{5.8}$ &$0.72$ & $0.70$\\
$[5.8]-[8.0]$ &$0.27$  & $0.25$ \\
\enddata
\tablecomments{\small{Mid-infrared flux ratios and colors for the
    nuclear region of the spiral galaxy
    NGC\,6872 determined from a fixed $3\farcs66$ circular aperture
    with $3\farcs66$,$8\farcs54$ inner, outer radius background
    annulus (Photometry) and from the color maps shown in Fig. 
    \protect\ref{fig:colormaps} (Map). Colors are given in
    magnitudes.}
}
\end{deluxetable}

\clearpage
\begin{deluxetable}{ccccccccc}
\tablewidth{0pc}
\tablecaption{X-ray Sources Associated with the Spiral Galaxy  NGC\,6872\label{tab:xraysrclist}}
\tablehead{\colhead{Label} & \colhead{J2000.0 Coordinates} & Total Counts$^a$ 
& $L_{\rm X}$ & H & M & S & $[H-M]$ & $[M-S]$ \\
  &  (RA, DEC) &  &($10^{39}$\ergs) & & & & &}
\startdata
 $1$  & $20\,\,17\,\,35.541,-70\,\,45\,\,06.55$ &$10.7 \pm 4.8$ &$0.64 \pm 0.21$& $ 8.2 \pm 4.3$ &$ 2.2 \pm 2.9$ & $0.3 \pm 2.3$ & $ 0.55$ & $ 0.17$ \\
 $2$  & $20\,\,17\,\,17.374,-70\,\,45\,\,07.76$ &$12.0 \pm 4.8$ &$0.59 \pm 0.21$& $ 7.0 \pm 4.0$ &$ 4.5 \pm 3.4$ & $0.4 \pm 2.3$ & $ 0.21$ & $ 0.34$\\
 $3$  & $20\,\,17\,\,00.954,-70\,\,46\,\,09.17$ &$42.2 \pm 7.6$ &$2.69 \pm 0.42$& $34.8 \pm 7.0$ &$ 6.7 \pm 3.8$ & $0.7 \pm 2.3$ & $ 0.67$ & $ 0.14$\\
 $4$  & $20\,\,16\,\,57.794,-70\,\,47\,\,03.59$ &$61.9 \pm 9.0$ &$3.61 \pm 0.44$& $29.5 \pm 6.5$ &$22.7 \pm 5.9$ &$ 9.8 \pm 4.3$ & $ 0.11$ & $ 0.21$\\
 $5$  & $20\,\,16\,\,56.795,-70\,\,46\,\,36.74$ &$83.3 \pm 10.2$ &$5.06 \pm 0.56$& $32.8 \pm 6.8$ &$38.8 \pm 7.3$ &$11.7 \pm 4.6$ & $-0.07$ & $ 0.33$\\
  N   & $20\,\,16\,\,56.542,-70\,\,46\,\,04.60$ &$140.4 \pm 13.6$ &$8.46 \pm 0.74$& $39.5 \pm 7.7$ &$61.9 \pm 9.3$ &$38.9 \pm 7.8$ & $-0.16$ & $ 0.16$\\
 $6$  & $20\,\,16\,\,55.666,-70\,\,46\,\,26.28$ &$30.3 \pm  6.8$ &$1.80 \pm 0.37$& $13.4 \pm 4.9$ &$15.1 \pm 5.1$ &$ 1.8 \pm 3.0$ & $-0.06$ & $ 0.44$\\
 $7$  & $20\,\,16\,\,53.386,-70\,\,46\,\,05.37$ &$25.9 \pm  6.7$ &$1.62 \pm 0.34$& $ 9.9 \pm 4.5$ &$15.1 \pm 5.3$ &$ 0.9 \pm 3.0$ & $-0.20$ & $ 0.55$ \\
 $8$  & $20\,\,16\,\,53.065,-70\,\,46\,\,38.67$ &$33.5 \pm  7.9$ &$1.70 \pm 0.29$& $ 4.3 \pm 4.4$ &$16.9 \pm 5.7$ &$12.4 \pm 5.1$ & $-0.38$ & $ 0.13$ \\
 $9$  & $20\,\,16\,\,52.363,-70\,\,46\,\,45.78$ &$9.1 \pm   4.3$ &$0.49 \pm 0.21$& $ 4.6 \pm 3.4$ &$ 4.7 \pm 3.4$ &$ 0.0 \pm 1.9$ & $-0.02$ & $ 0.52$\\
 $10$ & $20\,\,16\,\,52.293,-70\,\,46\,\,07.70$ &$13.7 \pm  5.4$ &$0.78 \pm 0.26$&$ 5.1 \pm 3.6$ &$ 6.4 \pm 4.0$ &$ 2.2 \pm 3.2$ & $-0.10$ & $ 0.30$\\
 $11$ & $20\,\,16\,\,49.125,-70\,\,46\,\,48.65$ &$43.7 \pm  7.8$ &$2.68 \pm 0.42$&$10.6 \pm 4.4$ &$29.3 \pm 6.6$ &$ 3.8 \pm 3.2$ & $-0.43$ & $ 0.59$\\
\enddata
\tablecomments{ {\small X-ray counts, luminosities, and colors for 
X-ray sources associated  with NGC\,6872. N denotes
NGC\,6872's nucleus. Columns are (1) source label; (2) J2000.0 WCS 
coordinates; (3)  $0.5-8$\,keV source counts (4) 
 $0.5-8$\,keV intrinsic X-ray luminosities 
 assuming a  $\Gamma = 1.6$ power law model with Galactic 
absorption $N_{\rm H} = 5 \times 10^{20}$\cms; (5, 6, 7) source counts
in the $2-8$\,keV (H), $1-2$\,keV (M), $0.5-1$\,keV (S) bands;
(8 and 9) X-ray colors defined as $[H-M] = (H-M)/(H+M+S)$
and $[M-S] = (M-S)/(H+M+S)$. $1\sigma$ uncertainties in the number of
counts are calculated using the Gehrels approximation for low count rate data.
}}
\end{deluxetable}

\begin{deluxetable}{ccccccc}
\tablewidth{0pc}
\tablecaption{Optical Properties of Star Clusters Coincident with
  X-ray Sources$^a$\label{tab:starclusters}}
\tablehead{\colhead{Source} & \colhead{M$_{\rm B}$} & 
\colhead{M$_{\rm V }$} & A$_{\rm B}$ & A$_{\rm V}$ & \colhead{Mass} & \colhead{Age}  
\\
 &  & & & &($10^5\,\Ms$) & (Myr) }
\startdata
$172$(X$4$) & $-14.03 \pm 0.01$ & $-13.81 \pm 0.02$ & $2.06$ &$1.55$ & $12$ &
$2.8$ \\
$221$(X$5$) & $-11.69 \pm 0.05$ & $-11.73 \pm 0.08$ & $ 0.66$ & $0.50$
& $5$ & $29$ \\
\enddata
\tablecomments{ {\small Col(1): star cluster label from Bastian \etal 
(2005) (coincident X-ray source from Table \protect\ref{tab:xraysrclist}); 
Col 2 (3): absolute B (V) magnitude, corrected for Galactic and internal
extinction; Col 4 (5): internal B (V) extinction in magnitudes; Col 6: star
cluster mass; Col 7: star cluster age. $^a$ All optical
properties are from Bastian \etal (2005) assuming a luminosity
distance of $55.5$\,Mpc (distance modulus $33.72$ ).
}} 
\end{deluxetable}

\begin{deluxetable}{cccc}
\tablewidth{0pc}
\tablecaption{X-ray to Optical Ratios for ULX Counterparts\label{tab:xor}}
\tablehead{\colhead{Source} & \colhead{XOR$_{\rm V}$} & 
\colhead{XOR$_{\rm B}$} & $\alpha ox$}
\startdata
$172$(X$4$) &$-0.93$ &$11.2$ &$1.7$  \\
$221$(X$5$) &$0.03$ &$13.9$ &$1.4$  \\
\enddata
\tablecomments{ {\small Col 1: Source label as in Table 
\protect\ref{tab:starclusters}; Col 2: V-band X-ray-to-optical ratio
(see eq. \protect\ref{eq:xorv}); Col 3: B-band X-ray-to-optical ratio (see
eq. \protect\ref{eq:xorb}); Col 4:  V-band to $1$\,keV two-point
spectral index, assuming ${\rm F}_{\nu} \propto \nu^{-\alpha ox}$.
}}
\end{deluxetable}

\end{document}